\documentclass[%
 reprint,
 amsmath,amssymb,
 aps,
]{revtex4-1}

\usepackage{graphicx}% Include figure files
\usepackage{dcolumn}% Align table columns on decimal point
\usepackage{bm}% bold math
\usepackage{color}

\begin{document}

\title{Phase-field theory of multi-component incompressible Cahn-Hilliard liquids}%ith Force line breaks with \\

\author{Gyula I. T\'oth}
\affiliation{Department of Physics and Technology, University of Bergen,\\All\'egaten 55, N-5007 Bergen, Norway}
\affiliation{Wigner Research Centre for Physics,\\P.O. Box 49, H-1525 Budapest, Hungary}
\email{Gyula.Toth@ift.uib.no}

\author{Mojdeh Zarifi}
\affiliation{Department of Physics and Technology, University of Bergen,\\All\'egaten 55, N-5007 Bergen, Norway}

\author{Bj\o rn Kvamme}
\affiliation{Department of Physics and Technology, University of Bergen,\\All\'egaten 55, N-5007 Bergen, Norway}


\date{\today}% It is always \today, today,

\begin{abstract}
In this paper a generalization of the Cahn-Hilliard theory of binary liquids is presented for multi-component incompressible liquid mixtures. First, a thermodynamically consistent convection-diffusion type dynamics is derived on the basis of the Lagrange multiplier formalism. Next, a generalization of the binary Cahn-Hilliard free energy functional is presented for arbitrary number of components, offering the utilization of independent pairwise equilibrium interfacial properties. We show that the equilibrium two-component interfaces minimize the functional, and demonstrate, that the energy penalization for multi-component states increases strictly monotonously as a function of the number of components being present. We validate the model via equilibrium contact angle calculations in ternary and quaternary (4-component) systems. Simulations addressing liquid flow assisted spinodal decomposition in these systems are also presented.
    
\end{abstract}

%\pacs{}% PACS, the Physics and Astronomy
\maketitle

%\tableofcontents

\section{Introduction}

Multi-component liquid mixtures are of continuously increasing scientific and industrial importance. For instance, it has recently been discovered, that controlled pattern formation in ternary colloidal emulsions / polymer mixtures could be used in producing advanced pharmaceutics, biochemical assays, or templating microporous materials \cite{ANGE:ANGE201406040,doi:10.1021/ma500302k}. Multi-component emulsions might also play important role in developing a new, efficient, and environmentally sound enhanced crude oil recovery technique \cite{ahearn,Iglauer201023,compareEOR,Song201493}. Although numerous theoretical studies addressing binary liquid flows are available, significantly less is known about ternary flows, and desperately less abouot 4 and more component systems. The continuum descriptions of binary systems undergoing phase separation originate from Cahn and Hilliard \cite{:/content/aip/journal/jcp/28/2/10.1063/1.1744102}, and was further improved by Cook \cite{Cook1970297} and Langer \cite{Langer197153,Langer19731649}. The binary theory was successfully extended also for ternary systems by de Fontaine \cite{DeFontaine1972297,Fontaine19731285}, Morral and Cahn \cite{Morral19711037}, Hoyt \cite{Hoyt19892489,Hoyt1990227}, and Maier-Paape et al \cite{stanislaus} (for many components), although the latter was applied exclusively for ternary systems. Coupling liquid flow to the Cahn-Hilliard theory is also possible on the basis of the Korteweg stress tensor \cite{Korteweg1901,doi:10.1080/00018737900101365} (also interpreted as the least action principle in statistical physics \cite{doi:10.1146/annurev.fl.20.010188.001301}), and has been done for binary systems by several authors \cite{Wheeler08081997,Anderson2000175,doi:10.1146/annurev.fluid.30.1.139}, thus resulting in a reasonable picture of binary liquids \cite{Tegze2005418}, while a liquid-flow coupled generalization of the Cahn-Hilliard model for arbitrary number of components was developed by Kim and Lowengrub \cite{KimLowengrub2005}, and later by Kim \cite{KimTernary}. The Kim-Lowengrub model was tested mainly for the ternary case, while quite limited calculations are available for 4-component systems. Furthermore, as it will be demonstrated in this paper, , the construction of neither the free energy functional nor the diffusion equations used by Kim and Lowengrub satisfy all conditions of physical and mathematical consistency, or if so, the constraints on the model parameters strongly limit the applicability of the theory. Therefore, the problem needs further investigation.
    
The main difficulty in describing many-component flows is finding appropriate extensions of both the thermodynamic functions and the dynamic properties for high-order multiple junctions. This is far from being trivial, mostly due to the lack of microscopic data. Nevertheless, one can extrapolate from the binary interfaces, while keeping physical and mathematical consistency. In case of spinodal decomposition, for example, physical consistency means, that the multi-component states of the material should be energetically less and less favorable with increasing number of components. Consequently, the system should converge to equilibrium configurations showing a single component -- binary interface -- trijunction topology. The conditions of mathematical consistency can be summarized as the \textit{condition of formal reducibility}, i.e. writing up the model for $N$ components, then setting the $N^{th}$ component to zero should result in the $N-1$ component model on the level of \textit{both} the free energy functional and the dynamic equations.

In this work, we formulate such a consistent generalization of the binary Cahn-Hilliard theory for arbitrary number of components, for which (i) the bulk states are absolute minima of the free energy functional, (ii) the two-component equilibrium interfaces represent stable equilibrium, and (iii) the energy of multiple junctions increases as a function of the number of components. In addition, the free energy density landscape has no multi-component local minima, therefore, the system cannot get trapped into a multi-component homogeneous state during spinodal decomposition.         Furthermore, a convection-diffusion dynamics is also developed, which (i) does not label the variables in principle, and (ii) extends / reduces naturally, when a component is added to / removed from the model. 

The paper is structured as follows. In Section II, we define first the relevant variables describing a multi-component liquid flow, together with introducing a general free energy functional formalism. Next, we study equilibrium via the Euler-Lagrange equations, and construct a general convection-diffusion dynamics. The application of the general framework for multi-component spinodal decomposition follows then in Section III. We construct a consistent extension of the  binary Cahn-Hilliard free energy functional for arbitrary number of components, and demonstrate both the physical and mathematical consistency of our approach. After presenting the numerical methods in Section IV, the validation of the model follows in Section V, including equilibrium contact angle measurements and modeling spinodal decomposition in both ternary and quaternary systems. The conclusions are summarized in Section VI.

\section{Theoretical framework}

\subsection{Energy functional formalism}

Consider a system of $N$ incompressible liquids of unique mass density $\rho$. In a mixture of the liquids, the \textit{mass fraction} of component $i$ reads $c_i=m_i/m$, where $m_i$ is the mass of component $i$ and $m=\sum_{i=1}^N m_i$ is the total mass in a control volume $V$. The mass fractions then sum up to 1 by definition, i.e. $\sum_{i=1}^N c_k =1$. Taking the limit $V \to 0$, all quantities become local, therefore, the (local and temporal), \textit{conserved} composition fields $c_i \to c_i(\mathbf{r},t)$ characterizing an inhomogeneous system can be introduced. The relation $\sum_{i=1}^N c_i=1$ transforms then into the following \textit{local} constraint:
\begin{equation}
\label{eq:local}
\sum_{i=1}^N c_i(\mathbf{r},t) = 1 \enskip .
\end{equation}
Assume that the Helmholtz free energy of the inhomogeneous non-equilibrium system can be expressed as a \textit{functional} of the fields:
\begin{equation}
F = \int dV \left\{ f[c_i(\mathbf{r},t),\nabla c_i(\mathbf{r},t)]\right\} \enskip ,
\end{equation}
where the integrand is a function of the fields and their gradients. This type of energy functional is called square gradient theory. In the literature the local constraint is often handled by eliminating one of the components already at the level of the free energy functional, thus resulting in an unconditional system. In contrast, Eq. (\ref{eq:local}) is taken into account here by using a Lagrange multiplier as:
\begin{equation}
\label{eq:cfdef}
\tilde{F}:=F-\int dV \left\{ \lambda(\mathbf{r},t)\left[ \sum_{i=1}^N c_i(\mathbf{r},t)-1\right] \right\} \enskip ,
\end{equation}
where $\tilde{F}$ is the \textit{conditional} free energy functional and $\lambda(\mathbf{r},t)$ the Lagrange multiplier. In our derivations, we will use this general formalism to derive consistent dynamic equations for the system.

\subsection{Equilibrium}

Equilibrium solutions represent extrema (minimum, maximum or saddle) of the free energy functional, therefore, they can be determined by solving the following Euler-Lagrange equations:  
\begin{equation}
\label{eq:elfinal}
\frac{\delta \tilde{F}}{\delta c_i} = \frac{\delta F}{\delta c_i}-\lambda(\mathbf{r}) = \tilde{\mu}_i^0 \enskip ,
\end{equation}     
where $\delta F/\delta c_i$ is the functional derivative of $F$ with respect to $c_i(\mathbf{r})$ ($i=1\dots N$), whereas $\tilde{\mu}_i^0=[(\delta F/\delta c_i) - \Lambda(\mathbf{r})]_{\mathbf{c}_0}$ is a diffusion potential belonging to a homogeneous reference state $\mathbf{c}_0=(c_1^0,c_2^0,\dots,c_N^0)$. Since the variables are conserved, the Lagrange multiplier cannot be expressed directly from Eq. (\ref{eq:elfinal}). Nevertheless, one can take the gradient of Eq. (\ref{eq:elfinal}) to eliminate the constant $\mu'_i$ [also containing the background value of $\lambda(\mathbf{r})$], yielding
\begin{equation}
\label{eq:gradel}
\nabla \frac{\delta F}{\delta c_i} = \nabla \lambda(\mathbf{r}) \enskip ,
\end{equation} 
or, equivalently
\begin{equation}
\label{eq:gradel2}
\nabla \left( \frac{\delta F}{\delta c_i} - \frac{\delta F}{\delta c_j} \right) = 0
\end{equation}
for any $(i,j)$ pairs. In general, $\nabla \lambda(\mathbf{r})$ can be eliminated from Eq. (\ref{eq:gradel}) as follows. Multiplying the equations by arbitrary weights $A_i \neq 0$, then summing them for $i=1\dots N$ results in: 
\begin{equation}
\label{eq:lelim}
\nabla\lambda(\mathbf{r}) = \sum_{i=1}^N a_i \nabla \frac{\delta F}{\delta c_i} \enskip ,
\end{equation}
where $a_i=A_i/\sum_{k=1}^N A_k \neq 0$ is a normalized coefficient, i.e. $\sum_{i=1}^N a_i=1$. Substituting Eq. (\ref{eq:lelim}) into Eq. (\ref{eq:gradel}), then re-writing the equations in a matrix form results in
\begin{equation}
\label{eq:matel}
(\mathbb{I}-\mathbf{e} \otimes \mathbf{a}) \cdot \nabla\frac{\delta F}{\delta \mathbf{c}} = 0 \enskip ,
\end{equation}    
where $\mathbb{I}$ is the $N \times N$ identity matrix, $\mathbf{e}=(1,1,\dots,1)^T$ is a column, while $\mathbf{a}=(a_1,a_2,\dots,a_N)$ a row vector, $\otimes$ denotes the dyadic (tensor or direct) product and $\delta F/\delta \mathbf{c}=(\delta F/\delta c_1,\delta F/\delta c_2,\dots,\delta F/\delta c_N)^T$ is the column vector of the functional derivatives. Note that the matrix $\mathbb{A}=\mathbb{I}-\mathbf{e} \otimes \mathbf{a}$ has a \textit{single} eigenvalue $s=0$ with eigenvector $\mathbf{e}$, thus prescribing equal functional derivative gradients in equilibrium, independently from the weights $\mathbf{a}$. (In other words, $\mathbf{e}$ is the algebraic representation of equilibrium.) Consequently, the solution of Eq. (\ref{eq:gradel}) coincides with the solution of Eq. (\ref{eq:gradel2}) for arbitrary positive $\{A_i\}$ weigths.

\subsection{Dynamic equations}

\subsubsection{Diffusion equations}

The incompressible multi-component flow is governed by an \textit{convection-diffusion} type dymamics. We start the derivation of the kinetic equations following Kim an Lowengrub \cite{KimLowengrub2005}. The diffusion equations follow from the mass balance for the individual components, thus resulting in \cite{KimLowengrub2005}:
\begin{equation}
\label{eq:dyn1}\rho\,\dot{c}_i = \nabla \cdot \mathbf{J}_i \enskip ,
\end{equation}
where $\dot{c}_i=\partial c_i/\partial t+\mathbf{v}\cdot\nabla c_i$ is the material derivative, $\mathbf{v}=\sum_{i=1}^N c_i \mathbf{v}_i$ is the mixture velocity, where $\mathbf{v}_i$ is the individual velocity field of the $i^{th}$ component. Furthermore, $\sum_i \mathbf{J}_i = 0$ applies for the diffusion fluxes, a condition emerging from $\sum_{i=1}^N c_i(\mathbf{r},t)=1 \to \sum_{i=1}^N \dot{c}_i(\mathbf{r},t)=0$. The diffusion fluxes can be then constructed as
\begin{equation}
\label{eq:chemcurr}
\mathbf{J}_i := \nu_i \nabla \tilde{\mu}_i
\end{equation}
(for example), where $\nu_i>0$ is the diffusion mobility of component $i$, and $\tilde{\mu}_i=\delta \tilde{F}/\delta c_i=\delta F/\delta c_i - \Lambda(\mathbf{r},t)$ is the generalized non-equilibrium chemical potential. Note that in equilibrium $\tilde{\mu}_i \to \tilde{\mu}_i^0$ (constant), thus indicating $\mathbf{J}_i=0$ and (consequently) $\dot{c}_i=0$. The Lagrange multiplier can be expressed as $\nabla\Lambda(\mathbf{r},t) = \sum_{i=1}^N \tilde{\nu}_i \nabla (\delta F/\delta c_i)$, where $\tilde{\nu}_i=\nu_i/\sum_{j=1}^N \nu_j > 0$. Using this in Eq. (\ref{eq:dyn1}), and introducing $\nu_i:=\kappa_i\Sigma$ (where $\Sigma=\sum_{k=1}^N \kappa_k$) results in
\begin{equation}
\label{eq:diffgen}
\mathbf{J}_i = \sum_{j=1}^N \kappa_{ij} \nabla \left( \frac{\delta F}{\delta c_i} - \frac{\delta F}{\delta c_j}  \right) \enskip ,
\end{equation}
where $\kappa_{ij}=\kappa_i\kappa_j$. Comparing Eq. (\ref{eq:diffgen}) and (\ref{eq:gradel2}), however, indicates $\mathbf{J}_i=0$ in equilibrium for \textit{arbitrary} $\kappa_{ij}$'s. The only condition for the mobilities emerges from the symmetry argument, that the variables should not be labeled, where labeling means that the time evolution of the system is not invariant under re-labeling the variables. The condition of no labeling yields \cite{PhysRevB.92.184105}
\begin{equation}
\label{eq:symmetry}
\kappa_{ij}=\kappa_{ji} \enskip ,
\end{equation}
in agreement with Onsager's approach of multi-component diffusion \cite{Onsager46}. In the Appendix of our recent study \cite{PhysRevB.92.184105} we pointed out that elimination of one of the variables by setting up $\mathbf{J}_i \propto (\delta F/\delta c_i) - (\delta F/\delta c_N)$ for $i=1\dots N-1$ labels the variables in principle, and contradicts to Onsager's reciprocal relations. The only exception is the fully symmetric system, i.e. when all interface thicknesses, interfacial tensions, and dynamic coefficients are equal. Note, that Eq. (\ref{eq:diffgen}) and Eq. (\ref{eq:symmetry}) offer a more general form for the constitutive equation than Eq. (\ref{eq:chemcurr}). In the latter we have only $N$ independent parameters, $\vec{\kappa}=(\kappa_1,\kappa_2,\dots,\kappa_N)$, and the mobility matrix $\mathbb{L}$ in the general form $\rho\,\dot{\mathbf{c}}=\nabla\cdot(\mathbb{L}\cdot \nabla\vec{\mu})$ emerge from these as $\mathbb{L}=\vec{\kappa} \otimes \vec{\kappa}$. In contrast, according to Eq. (\ref{eq:diffgen}) and (\ref{eq:symmetry}), we may chose $N(N-1)/2$ free parameters $\{\kappa_{ij}\}$ in general, and the elements of the mobility matrix are calculated as $L_{ii}=\sum_{j\neq i}\kappa_{ij}$, and $L_{ij}=-\kappa_{ij}$ for $i \neq j$. Although Eq. (\ref{eq:chemcurr}) and (\ref{eq:diffgen}) coincide in equilibrium, the general construction becomes significant for $N \geq 4$, where the number pairs are greater than $N$.

The remaining issue which has to be considered is the condition of "formal reducibility" for the dynamic equations. An elegant solution of the problem introducing mobility matrices on geometric basis was published by Bollada, Jimack and Mullis \cite{Bollada2012816}. The authors proposed symmetric mobility matrices reducing formally. For example, in case of $\kappa_{ij}(c_i,c_j)=[c_i/(1-c_i][c_j/(1-c_j)]$ the $k^{th}$ row and column of the mobility matrix vanish, and the mobility matrix of an $N-1$-component system is recovered. Note, however, that such a mobility matrix can be "dangerous" with respect to the free energy functional, meaning that non-equilibrium states may become stationary, since the equality of the functional derivative gradients is not a necessary condition for a stationary solution. Speaking mathematically more precisely, the eigenvalue $s=0$ (representing stationary solution) of the mobility matrix $\mathbb{L}$ has multiplicity greater than 1 in case of at least 1 vanishing field. The components of the corresponding eigenvectors are equal at the positions of non-vanishing fields, otherwise they are arbitrary. \textit{Therefore, a stationary state of the dynamics does not necessarily represent equilibrium solution}. Nevertheless, as discussed in our recent paper \cite{PhysRevB.92.184105}, if one can prove that the $n+m$-component natural extensions of all equilibrium solutions emerging from the $n$-component model also represent equilibrium in the $n+m$-component model for any $n,m\geq 1$, then the Bollada-Jimack-Mullis matrix is not dangerous with respect to the free energy functional. Having such a functional, although being necessary, is not satisfactory, since the dynamics must satisfy also the \textit{second law of thermodynamics}, i.e. the entropy production rate must be non-negative. This requirement can be addressed by considering the Kim-Lowengrub model in the constant density limit. The condition for the contribution of the diffusion equations to the entropy production rate reads \cite{KimLowengrub2005}:
\begin{equation}
\label{eq:entropy}
\sum_{i=1}^N \nabla\hat{\mu}_i \cdot \mathbf{J}_i \geq 0 \enskip .
\end{equation}
Here $\hat{\mu}_i = (\delta F/\delta c_i)+p$, where $p$ is the non-equilibrium thermodynamic pressure. According to Eq. (\ref{eq:diffgen}), $\mathbf{J}_i=\sum_{j=1}^N L_{ij}\nabla(\delta F/\delta c_j)$, where $\sum_{j=1}^N L_{ij}=0$, therefore, Eq. (\ref{eq:entropy}) results in
\begin{equation}
\sum_{i,j}L_{ij}\left(\nabla\frac{\delta F}{\delta c_i}\cdot\nabla\frac{\delta F}{\delta c_j}\right) \geq 0 \enskip , 
\end{equation}
thus indicating that the mobility matrix $\mathbb{L}$ must be \textit{positive semi-definite}. Therefore, the original Bollada-Jimack-Mullis matrix is modified as 
\begin{equation}
\label{eq:BJMmobmat}
\kappa_{ij}(c_i,c_j) := \kappa_{ij}^0 \left|\frac{c_i}{1-c_i}\right|\left|\frac{c_j}{1-c_j}\right| \enskip ,
\end{equation}   
where $\kappa_{ij}^0>0$'s are arbitrary constants. The absolute value is necessary for a simple reason: The solution may slightly leave the physical regime ($0 \leq c_i \leq 1$ for $i=1\dots N$) in the simulations because of numerical reasons. Nevertheless, small perturbations around \textit{stable} equilibrium solutions relax naturally for a positive semi-definite mobility matrix without any further artificial modifications, such as overwriting the solution. This should be true for at least the bulk components and the binary equilibrium interfaces. The positive semi-definiteness of this matrix has been verified numerically case by case for the particular matrices we used in our calculations and simulations.

\subsubsection{Navier-Stokes equation}

The velocity field is governed by the following Navier-Stokes equation (emerging from the momentum balance for the components) \cite{KimLowengrub2005}:
\begin{equation}
\label{eq:dyn2}\rho\,\dot{\mathbf{v}} = \nabla \cdot ( \mathbb{R} + \mathbb{D} ) \enskip ,
\end{equation}
where $\mathbb{R}$ and $\mathbb{D}$ are the reversible and irreversible stresses, respectively. The viscous stress of a multi-component Newtonian liquid can be approximated as:
\begin{equation}
\label{eq:dissipative}
\mathbb{D} = \eta [\left(\nabla \otimes \mathbf{v}) + (\nabla \otimes \mathbf{v})^T \right] \enskip ,
\end{equation}   
where $\eta=\sum_{i=1}^N c_i \eta_i$ is the local shear viscositiy, calculated from the viscosities of the bulk components, $\eta_i$. Furthermore, the reversible stress has the general Korteweg form \cite{Korteweg1901,doi:10.1080/00018737900101365}:
\begin{equation}
\label{eq:reversible}
\mathbb{R} = -p \,\mathbb{I} + \mathbb{A} \enskip ,
\end{equation} 
where $\tilde{p}$ is a non-equilibrium generalization of the equilibrium thermodynamic pressure:
\begin{equation}
\label{eq:pressure}
-p = \tilde{f} - \sum_{i=1}^N c_i \frac{\delta \tilde{F}}{\delta c_i} = - \tilde{p} + \Lambda(\mathbf{r},t) \enskip,
\end{equation}
where $\tilde{f}$ is the integrand of $\tilde{F}$ defined by Eq. (\ref{eq:cfdef}), and $-\tilde{p}=f - \sum_{x=1}^N c_i (\delta F/\delta c_i)$. Furthermore, $\mathbb{A}$ is a general non-diagonal tensor, which can be determined from the condition of mechanical equilibrium, often formulated as a generalized Gibbs-Duhem relation \cite{doi:10.1146/annurev.fl.20.010188.001301,Wheeler08081997,Anderson2000175}
\begin{equation}
\label{eq:gibbsD}
\nabla \cdot \mathbb{R} = - \sum_{i=1}^N c_i \nabla \frac{\delta \tilde{F}}{\delta c_i} \enskip .
\end{equation} 
Using Eq. (\ref{eq:reversible}) in (\ref{eq:gibbsD}) then yields
\begin{equation}
\label{eq:nondiag}
\mathbb{A} = -\sum_{i=1}^N \left( \nabla c_i \otimes \frac{\partial f}{\partial \nabla c_i} \right) \enskip ,
\end{equation}
showing that the flow operator does not contain the Lagrange multiplier. This result is in agreement with previous results \cite{KimLowengrub2005}. Furthermore, since the liquid mixture is incompressible and all the components have the same density, we also have the condition
\begin{equation}
\label{eq:incomp}
\nabla\cdot\mathbf{v} = 0 \enskip .
\end{equation}
Although this condition results in a \textit{degeneracy} in the velocity field, it is resolved by the Lagrange multiplier $\Lambda(\mathbf{r},t)$ in Eq. (\ref{eq:pressure}).

\section{Multi-component Cahn-Hilliard liquid}

\subsection{Free energy functional}

The free energy functional of a general, multi-component Cahn-Hilliard liquid is formulated as:
\begin{equation}
\label{eq:CHfunc}
F = \int dV \left\{ f(\mathbf{c}) + \frac{\epsilon^2(\mathbf{c})}{2}\sum_{i=1}^N (\nabla c_i)^2\right\} \enskip ,
\end{equation}
where the \textit{multi-well} free energy landscape $f(\mathbf{c})$ is constructed as \cite{PhysRevB.92.184105}:
\begin{equation}
\label{eq:energysurf}
f(\mathbf{c}) := w(\mathbf{c})\,g(\mathbf{c}) + A_3 f_3(\mathbf{c}) \enskip ,
\end{equation}
where
\begin{equation}
\label{eq:multiwell}
g(\mathbf{c}) = \frac{1}{12} + \sum_{i=1}^N \left( \frac{c_i^4}{4} - \frac{c_i^3}{3} \right) + \frac{1}{2} \sum_{i<j} c_i^2 c_j^2 \enskip .
\end{equation}
In Eq. (\ref{eq:multiwell}), the double sum stands for a summation for all pairs, i.e. $\sum_{i<j}=\sum_{i=1}^{N-1}\sum_{j=i+1}^N$. Following Kazaryan \cite{PhysRevB.61.14275}, the coefficients $w(\mathbf{c})$ and $\epsilon^2(\mathbf{c})$ interpolating between the component pairs read as:
\begin{equation}
w(\mathbf{c}) = \frac{\sum_{i<j} w_{ij} c_i^2 c_j^2}{\sum_{i<j} c_i^2 c_j^2} \quad \text{and} \quad \epsilon^2(\mathbf{c}) = \frac{\sum_{i<j} \epsilon^2_{ij} c_i^2 c_j^2}{\sum_{i<j} c_i^2 c_j^2} \enskip .
\end{equation}
Furthermore, the "triplet" term is defined as:
\begin{equation}
\label{eq:triplet}
f_3(\mathbf{c}) := \sum_{i<j<k} |c_i|\,|c_j|\,|c_k| \enskip , 
\end{equation}
where the sum is for all different $(i,j,k)$ triplets, i.e. $i\neq j$, $i\neq k$, and $j \neq k$, $i,j,k=1\dots N$. The usual (Gibbs-simplex) representation of the free energy landscape is shown in Fig. 1(a)-(d) for symmetric and asymmetric ternary systems, in case of $A_3=0$ and $A_3 \neq 0$, respectively. We note, that similar terms are used by some authors \cite{kumarbritta,PhysRevB.91.174109} to control the presence of the third component at binary interfaces, however, our approach is quite different than theirs, as it will be shown.

\subsection{Interfaces, energy hierarchy and stability}

When exactly two components are present, i.e. $c_i(\mathbf{r})+c_j(\mathbf{r})=1$ for $i \neq j$, and $c_k=0$ for all $k \neq i,j$, Eq. (\ref{eq:CHfunc}) reduces to the usual binary Cahn-Hilliard free energy functional:
\begin{equation}
\label{eq:CHfuncpair}
  F_{ij} = \int dV \left\{ w_{ij} [c(1-c)]^2 + \epsilon_{ij}^2 (\nabla c)^2 \right\} \enskip ,
\end{equation}
therefore, $\epsilon^2_{ij}$'s and $w_{ij}$'s can be related to the interfacial tension ($\sigma_{ij}$) and interface thickness ($\delta_{ij}$) as:
\begin{equation}
  w_{ij} = 3 (\sigma_{ij}/\delta_{ij}) \quad \text{and} \quad \epsilon_{ij}^2 = 3( \sigma_{ij} \delta_{ij}) \enskip ,
\end{equation} 
where the interface thickness is defined by the binary equilibrium interface solution 
\begin{equation}
\label{eq:equilibriumintf}
c_{ij}(x)=\{1+\tanh[x/(2 \,\delta_{ij})]\}/2 \enskip,
\end{equation} 
while the interfacial tension reads
\begin{equation}
\sigma_{ij}=\int_{-\infty}^{+\infty}dx \left\{ w_{ij} [c_{ij}(x)]^2[1-c_{ij}(x)]^2 + \epsilon_{ij}^2 [\partial_x c_{ij}(x)]^2 \right\} \enskip .
\end{equation}
The general functional defined by Eq. (\ref{eq:CHfunc}) has two practical features: 
\begin{itemize}
\item $F$, \textit{together with} $\delta F/\delta c_i$ reduce formally, i.e. writing up $F$ (and $\delta F/\delta c_i$) for $N$ fields, then applying $c_N\equiv 0$ results in the expressions derived directly in the $N-1$-component model. This, together with Eq. (\ref{eq:BJMmobmat}) result in the formal reducibility of the dynamic equations too;

\item All two-component equilibrium interfaces $c_{kl}(x)=[\{1+\tanh[x/(2\,\delta_{kl})]\}/2$ represent equilibrium in the \textit{complete}, $N$-component model. In other words, \textit{the binary planar interfaces represent equilibrium in the $N$-component system} (see Appendix A for details).
\end{itemize}
We mention, that the latter does not apply for almost any of previous multiphase/multicomponent descriptions \cite{PhysRevB.92.184105}. Nevertheless, it is an essential feature because of the followings: Eq. (\ref{eq:equilibriumintf}) represents only a \textit{conditional} extremum, since it is calculated in the $c_i(\mathbf{r})+c_j(\mathbf{r})=1$ binary subspace. Therefore, there's no guarantee that it is also a solution of the complete variational problem defined by Eq (\ref{eq:gradel2}). In case of several existing multiphase descriptions the case is indeed this, the equilibrium two-component interfaces do not represent equilbrium of the general, $N$-component model, doe to the inconsistent generalization of the free energy functional. The problem is resolved on various ways, including the introduction of non-variational dynamics / degenerated mobility matrix, or penalizing free energy terms for ternary states, as also discussed in details in our recent work \cite{PhysRevB.92.184105}. In contrast, our description is totally free of these artificial modifications.

\begin{figure}[t]
\includegraphics[width=0.49\linewidth]{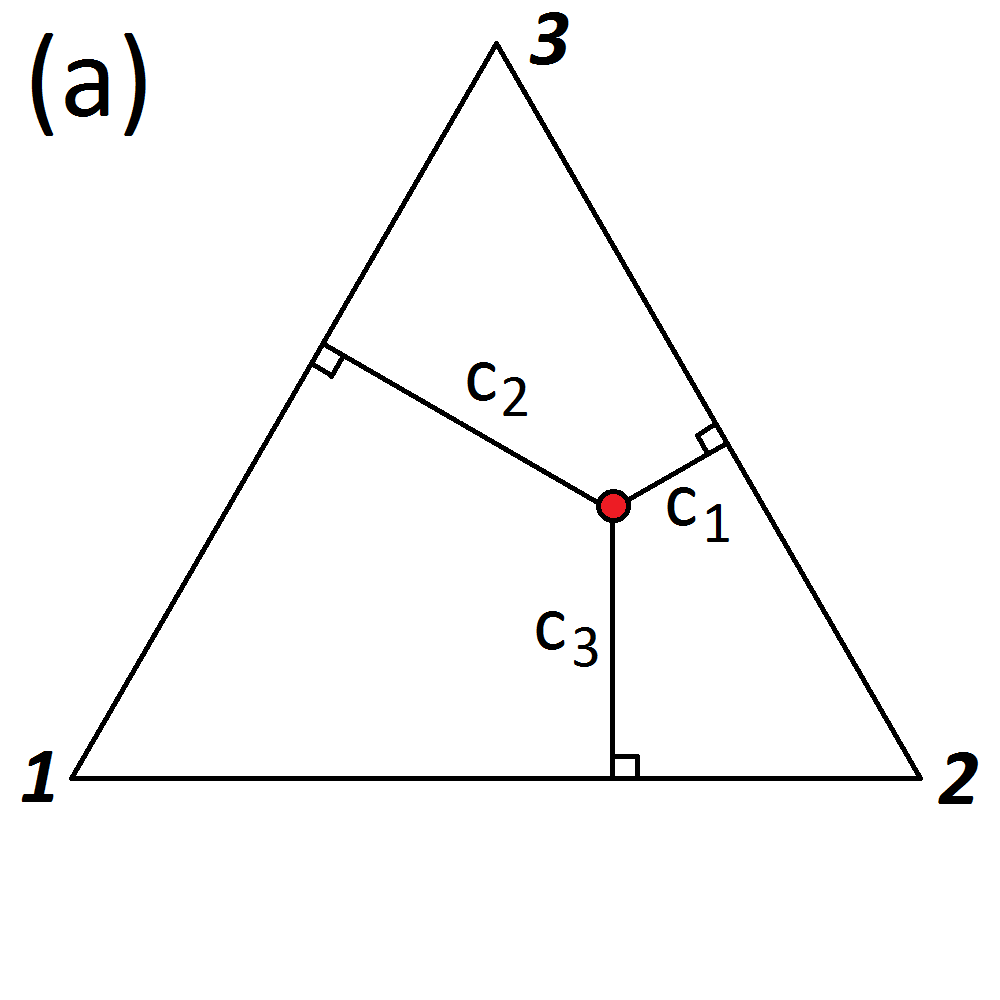}
\includegraphics[width=0.49\linewidth]{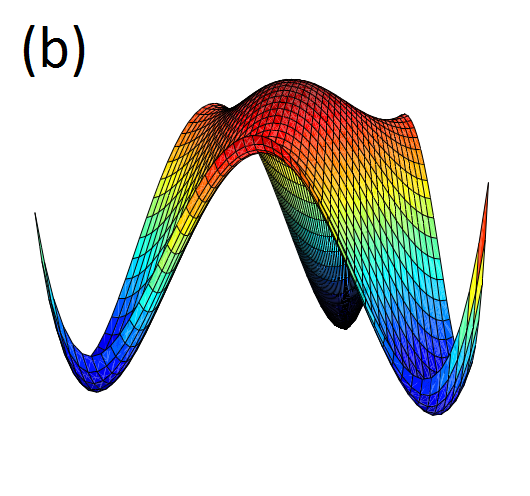}\\
\includegraphics[width=0.49\linewidth]{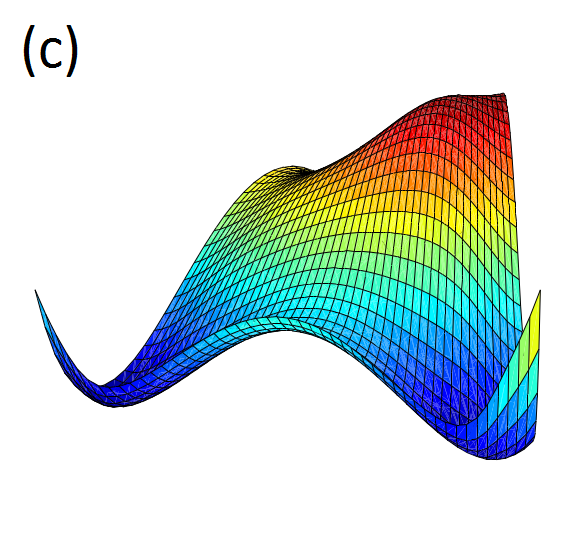}
\includegraphics[width=0.49\linewidth]{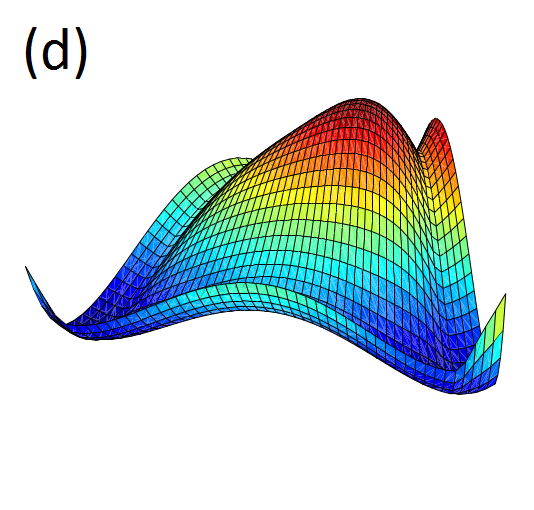}\\
\caption{Gibbs simplex and free energy landscapes $f(\mathbf{c})$ in ternary systems. (a) Gibbs simplex in a ternary system. The compositions in the red dot are measured perpendicular to the edges of the triangle. If all the edges measure 1 unit, $c_1+c_2+c_3=1$. The vertices (denoted by bold numbers) correspond to bulk components, i.e. $c_i=1$ at vertex $i$, where $i=1,2,3$. (b) Free energy density in the symmetric system without triplet term (i.e. $A_3=0$). (c-d) Free energy landscapes in an asymmetric ternary system ($w_{12}=1.5\,w_0$, $w_{13}=1.0\,w_0$, $w_{23}=0.5\,w_0$) in case of $A_3=0$ (panel c) and $A_3=1.0\,w_0$ (panel d). The minima of the free energy landscapes correspond to the vertices of the Gibbs simplex displayed in panel a.}
\end{figure}
In a symmetric system ($\epsilon_{ij}^2\equiv \epsilon^2_0$ and $w_{ij}\equiv w_0$) without triplet energy contribution ($A_3=0$), Eq. (\ref{eq:energysurf}) is a finite-degree polynomial penalizing the multi-component states as follows:
\begin{equation}
\label{eq:punish}
f(\mathbf{c}_n) = \frac{1}{12} \left( 1 - \frac{1}{n^2}\right) \enskip ,
\end{equation}  
where $\mathbf{c}_n=\mathcal{P}[\{1/n,1/n,\dots,1/n,0,0,\dots,0\}]$. Here $\mathcal{P}[.]$ stands for an arbitrary permutation of the components of the vector argument $\{c_1,c_2,\dots,c_N\}$, where $n$ elements have the value $1/n$ and all the others are $0$, while $n=1\dots N$. Eq. (\ref{eq:energysurf}) then penalizes equally the $n$-component states, and the energy increases strictly monotonously as a function of the number of components being present. This feature also applies for \textit{arbitrary} $A_3 \geq 0$ in the symmetric system for the triplet term defined by Eq. (\ref{eq:triplet}) (see Appendix B for the derivation).

\begin{figure}[t]
\includegraphics[width=0.49\linewidth]{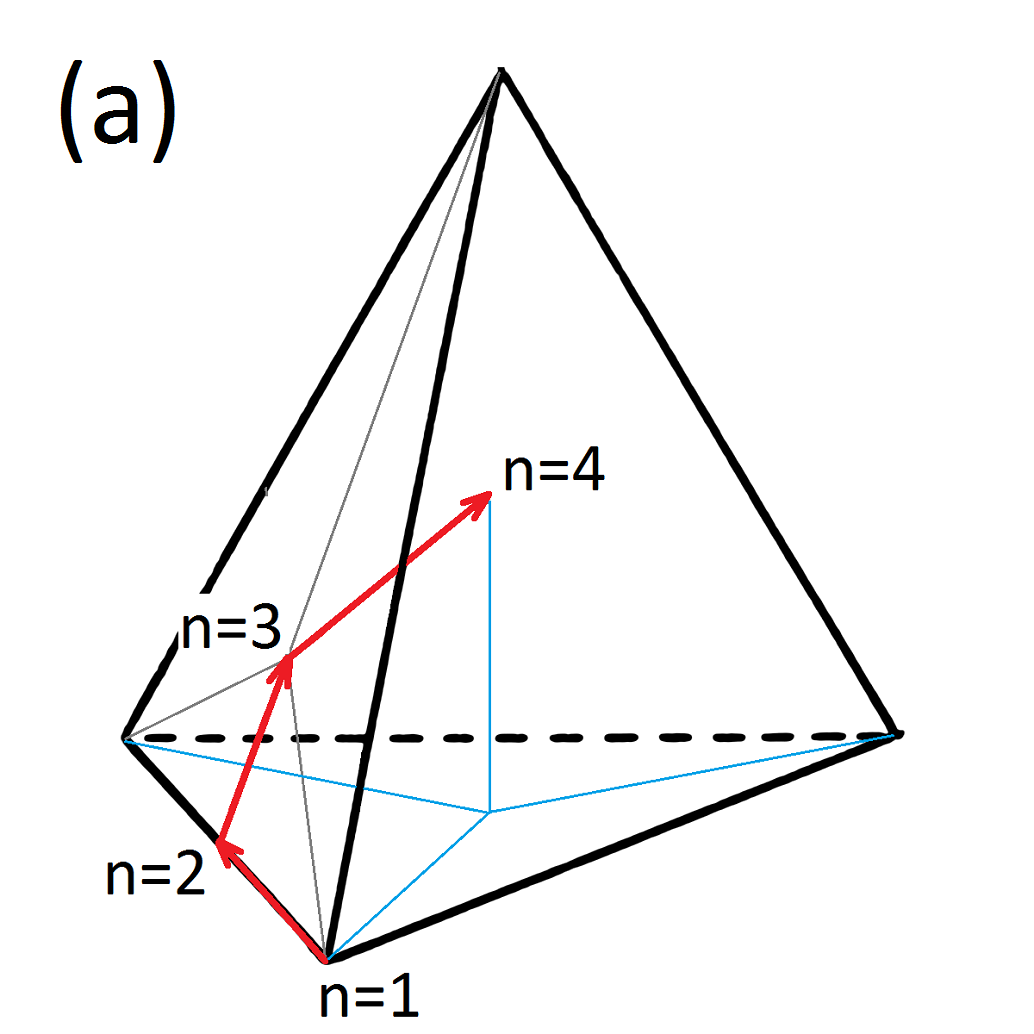}
\includegraphics[width=0.49\linewidth]{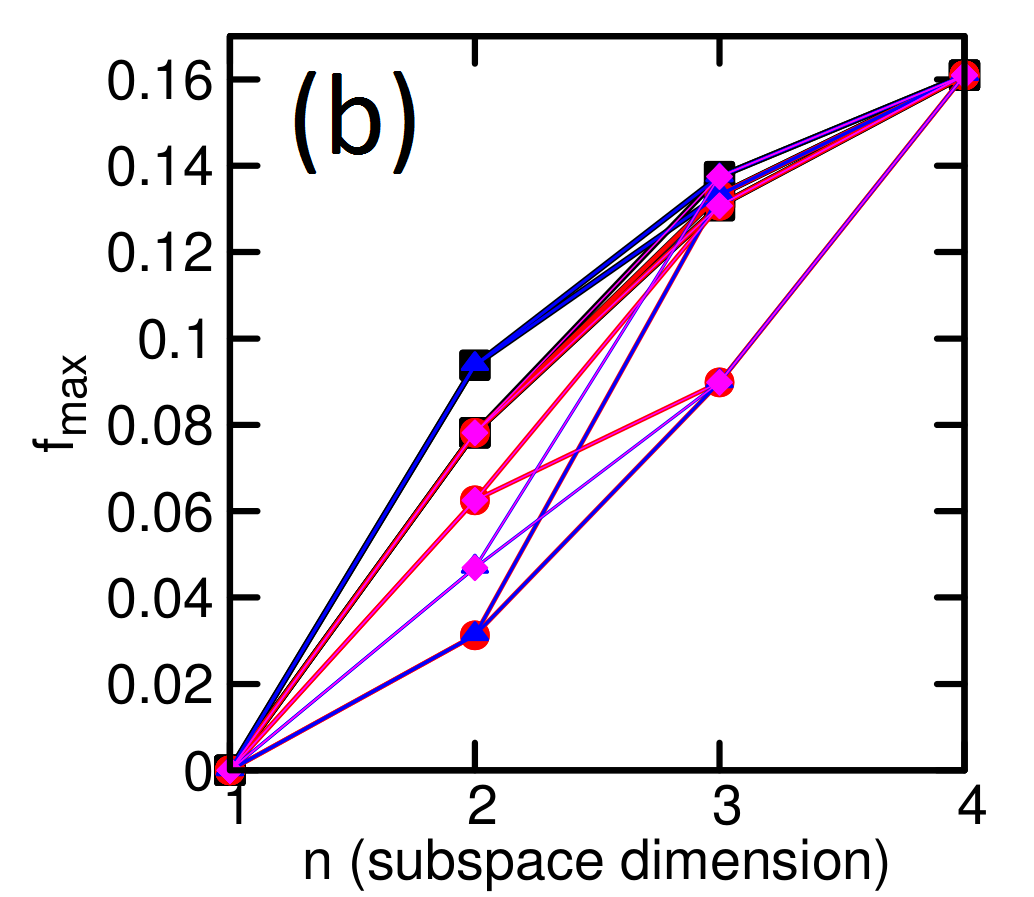}
\caption{Degeneracy of the subspace extrema in an asymmetric quaternary system ($w_{12}=1.25\,w_0$, $w_{13}=1.5\,w_0$, $w_{23}=0.5\,w_0$, $w_{14}=1.25\,w_0$, $w_{24}=1.0\,w_0$, $w_{34}=0.75\,w_0$), for $A_3=1.0\,w_0$. (a) A possible path stars in a vertex ($n=1$) representing the absolute minimum of the free energy density, then passes the location of a binary ($n=2$) and a ternary ($n=3$) maximum, while finally arrives at the location of the single quaternary ($n=4$) maximum, which is the absolute maximum of the free energy density. (b) Sequences of subspace extrema along all possible paths illustrated in panel a.}
\end{figure}

Interestingly, the strictly monotonous tendency of the subspace extrema seems to be valid even for asymmetric systems, however, both $f(\mathbf{c}_n)$ and $\mathbf{c}_n$ have now degeneracy, since both the location and the value of the subspace maxima can be different. This is illustrated in Fig. 2, which shows the degenerated hierarchy of the subspace extrema in case of asymmetry for $N=4$. Since the $n=2$ and $3$-component subspace maxima of the Gibbs simplex now can be different, one can define a "path" on the Gibbs simplex as follows. A path starts in a vertex (representing a bulk component), then jumps to the location of the maximum of one of the connecting edges [denoted by $n=1$ and $n=2$ in Fig. 2(a), respectively]. From here, we jump to the location of the maximum of one of the connecting planes ($n=3$), while the final point is the location of the global maximum inside the tetrahedron. Fig. 2(b) shows the energy density in the subspace maxima (symbols) along all possible bulk $\to$ binary $\to$ ternary $\to$ quaternary paths (denoted by the connecting lines). It seems that all the 24 possible paths prescribe strictly monotonously increasing energy sequence. If the free energy landscape does not have any other extrema, and all the extrema except the vertices represent maxima, then this behavior, together with the fact, that the free energy functional penalizes any spatial variation of the fields, suggest, that an $N$-component system undergoes spinodal decomposition, and without becoming trapped into a high-order state, i.e., the system never prefers high order multiple junctions, independently from the number of components.

\begin{figure}[t]
\begin{center}
\includegraphics[width=1.0\linewidth]{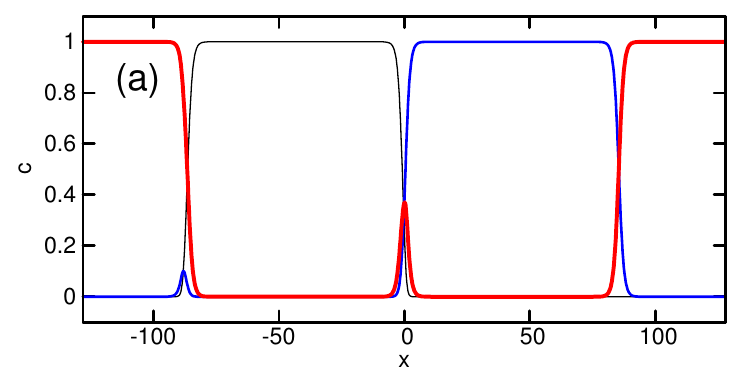}
\includegraphics[width=1.0\linewidth]{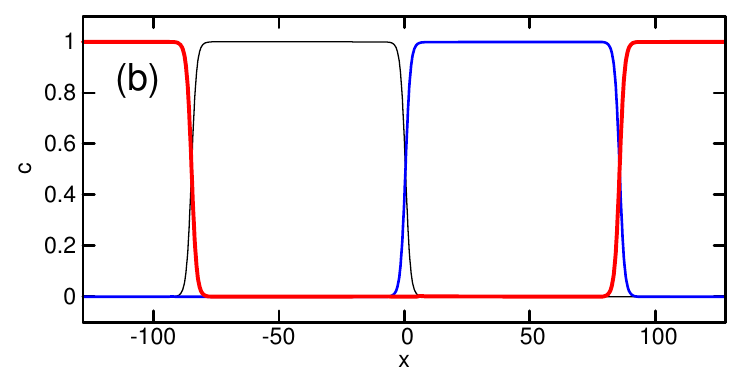}
\caption{Two-component equilibrium interfaces in an asymmetric ternary system ($w_{12}=1.5\,w_0$, $w_{13}=1.0\,w_0$, $w_{23}=0.5\,w_0$) in case of (a) $A_3=0$ and (b) $A_3=1.0\,w_0$. Note that, in case of $A_3=0$, $c_3$ (thick red) and $c_2$ (normal blue) appear on the $(1,2)$ and $(1,3)$ interfaces, respectively, while $c_1$ (thin black) does not appear on the $(2,3)$ interface, which has the lowest energy. Applying the triplet term then prevents the appearance of the third component at any two-component interfaces.}
\end{center}
\end{figure}

Although we constructed a free energy functional, which is expected to result in spinodal decomposition for an energy minimizing dynamics, and for which the binary planar interfaces together with the bulk states are equilibrium solutions, the interfaces may become unstable in case of asymmetry for $A_3=0$. The reason is, that the $A_3=0$ free energy landscape is "weak" for the multi-component states, meaning that the energy increases too slowly as a function of $n$: the energy difference between $f(1,0,\dots)$ and $f(1/2,1/2,0,\dots)$ is much more significant than that of between $f(1/2,1/2,0,\dots)$ and $f(1/3,1/3,1/3,0,\dots)$ [and so on, see Fig. 1(b) and Eq. (\ref{eq:punish})]. This means, that in case of asymmetry [see Fig. 1(c)], the shift in the location of three-component maximum can be significant, and therefore it can destabilize the binary planar interface on the closest edge (or, as a matter of fact, on any other edges, except the one with the lowest interfacial tension). To stabilize the (otherwise equilibrium) binary planar interfaces, we apply the triplet term described by Eq. (\ref{eq:triplet}). Choosing a sufficiently large amplitude $A_3$ shifts the location and increases the value of the ternary maximum of the free energy landscape [see Fig 1(d)], thus resulting in the re-stabilization  of the interfaces. The phenomenon is also illustrated in Fig 3. The figure shows the numerical solution of the 1-dimensional Euler-Lagrange problem in an asymmetric ternary system for $A_3=0$ [panel (a)] and $A_3\neq 0$ [panel (b]). We used finite difference method with explicit time stepping to solve the Euler-Lagrange problem $\nabla(\delta F/\delta c_i)=\nabla(\delta F/\delta c_j)$, together with periodic boundary conditions. As one can see, the third component appears at both the $(1,2)$ and $(1,3)$ interfaces in case of $A_3=0$ [see panel (a)], showing that the free energy landscape is weak with respect to the gradient term, and the binary planar interfaces, although representing equilibrium, are not stable. The only stable interface is the $(2,3)$ interface, which has the lowest energy. Nevertheless, choosing $A_3=1$ solves the problem [see panel (b)], since as the three-component maximum of the free energy landscape increases, the interfaces become stable.

Summarizing, Eq. (\ref{eq:CHfunc}) prescribes a multi-component free energy functional, which results in stable bulk states and binary interfaces in equilibrium, even for asymmetric systems, while high-order multiple states are penalized increasingly as a function of the components. This behavior results in spinodal decomposition in a system of arbitrary number of components. Therefore, Eq. (\ref{eq:CHfunc}) is a suitable generalization of the binary Cahn-Hilliard free energy functional. The triplet term $f_3(\mathbf{c})$ has no effect on the bulk ($n=1$) and binary states ($n=2$), and on the structure and hierarchy of the subspace extrema of the free energy landscape, but controls the energy of multi-component (ternary and up) states. Therefore, it is an ideal tool to control the \textit{stability} of the binary planar interfaces.

\subsection{Parameters and scaling}

To anchor the mobilities $\kappa_{ij}^0$ in Eq. (\ref{eq:BJMmobmat}) to measurable quantites, we first take Eq. (\ref{eq:dyn1}) in the binary limit $c_i=u$, $c_j=1-u$, and $c_k=0$ for $i \neq j$ and $k\neq i,j$. In case of $\mathbf{v}=0$ it yields
\begin{equation}
\label{eq:difflimit}
\rho \frac{\partial u}{\partial t} = \kappa_{ij}^0 \nabla \frac{\delta F}{\delta u} \enskip ,
\end{equation} 
and $\partial c_k/\partial t = 0$ for $k \neq i,j$. The functional derivative reads  $\delta F/\delta u = 2 \left\{ w_{ij} [u(1-u)(1-2 u)]-\epsilon_{ij}^2 \nabla^2 u \right\}$. For $u=\delta u \to 0$, Eq. (\ref{eq:difflimit}) becomes $\rho (\partial_t \delta u) = 2 \kappa_{ij}^0 w_{ij} (\nabla^2 \delta u)$, yielding thus the diffusion constant $D_{ij}=(2\kappa_{ij}^0 w_{ij})/\rho$ of the $i^{th}$ component in the bulk $j^{th}$ component. The mobility is then related to the diffusion constant via 
\begin{equation}
\frac{2 w_{ij}\kappa_{ij}^0}{D_{ij}} = \frac{2 w_{ij}\kappa_{ij}^0}{D_{ji}} = \rho \enskip ,
\end{equation}
where the second equation emerges from the symmetry of $\kappa_{ij}^0$. Therefore, the diffusion constant of the $j^{th}$ component in the $i^{th}$ one the same as that of the $i^{th}$ component in the $j^{th}$ one in our approach. Scaling the length as $\mathbf{r}:=\lambda\hat{\mathbf{r}}$, and introducing $D_{ij}:=D_0\hat{D}_{ij}$ yields the time scale $\tau=\lambda^2/D_0$ in $t:=\tau\hat{t}$, while using $w_{ij}:=w_0\hat{w}_{ij}$, and $\epsilon^2_{ij}:=\epsilon_0^2 \hat{\epsilon}_{ij}^2$ result in the dimensionless diffusion equations
\begin{equation}
\label{eq:dyns1}
\frac{d c_i}{d\hat{t}} = \hat{\nabla} \cdot \hat{\mathbf{J}}_i \enskip .
\end{equation}
The dimensionless diffusion fluxes read
\begin{eqnarray}
\label{eq:scaledJ}\hat{\mathbf{J}}_i &=& \sum_{j=1}^N \hat{\kappa}^0_{ij} \, h(c_i,c_j) \, \hat{\nabla} \left( \frac{\delta \hat{F}}{\delta c_i} - \frac{\delta \hat{F}}{\delta c_j} \right) \\
\label{eq:scaledFD}\frac{\delta \hat{F}}{\delta c_i} &=& \frac{\partial (\hat{w}\,g+\hat{A}_3\,f_3)}{\partial c_i} + \frac{\delta_0^2}{\lambda^2} \left[ \frac{\partial \hat{\epsilon}^2}{\partial c_i} (\hat{\nabla}\mathbf{c})^2  - \hat{\epsilon}^2 \hat{\nabla}^2 c_i \right] \enskip ,
\end{eqnarray}
where $\delta_0^2=\epsilon_0^2/w_0$. Furthermore, $h(c_i,c_j)=|c_i/(1-c_i)||c_j/(1-c_j)|$ and
\begin{equation}
2\,\hat{\kappa}^0_{ij}=\hat{D}_{ij}/\hat{w}_{ij} \enskip .
\end{equation}
The dimensionless coefficients read as
\begin{equation}
\hat{w}=\frac{\sum_{i<j} \hat{w}_{ij} c_i^2 c_j^2 }{\sum_{i<j} c_i^2 c_j^2} \quad \text{and} \quad \hat{\epsilon}^2=\frac{\sum_{i<j} \hat{\epsilon}^2_{ij} c_i^2 c_j^2} {\sum_{i<j}c_i^2 c_j^2} \enskip ,
\end{equation}
while $\hat{A}_3=A_3/w_0$. Introducing the dimensionless interfacial tensions $\sigma_{ij}:=\sigma_0\hat{\sigma}_{ij}$ and interface thicknesses $\delta_{ij}:=\delta_0\hat{\delta}_{ij}$, and considering $\epsilon^2_{ij}=3(\sigma_{ij}\delta_{ij})$ and $w_{ij}=3(\sigma_{ij}/\delta_{ij})$ yield the scales
\begin{equation}
\epsilon_0^2=3(\sigma_0\delta_0) \quad \text{and} \quad w_0=3(\sigma_0/\delta_0) \enskip ,
\end{equation} 
and
\begin{equation}
\hat{\epsilon}^2_{ij}=\hat{\sigma}_{ij} \hat{\delta}_{ij} \quad \text{and} \quad \hat{w}_{ij}=\hat{\sigma}_{ij}/\hat{\delta}_{ij} \enskip . 
\end{equation}
Furthermore, $\epsilon_0^2/w_0=\delta_0^2$ in Eq. (\ref{eq:scaledFD}). The dimensionless Navier-Stokes equation reads:
\begin{equation}
\label{eq:dyns2}
\frac{d\hat{\mathbf{v}}}{d\hat{t}}=\hat{\nabla}\cdot\hat{\mathbb{P}} \enskip ,
\end{equation} 
where
\begin{equation}
\label{eq:scaledP}
\hat{\mathbb{P}}=\hat{a}\,\hat{\mathbb{A}}(\mathbf{c})+\hat{\eta} \,\hat{\mathbb{D}}(\hat{\mathbf{v}}) \enskip .
\end{equation}
Here the dimensionless flow field generator $\hat{\mathbb{A}}(\mathbf{c})$ and the viscous stress $\hat{\mathbb{D}}(\hat{\mathbf{v}})$ read:
\begin{eqnarray}
\label{eq:scaledR}\hat{\mathbb{A}}(\mathbf{c}) &=& -\hat{\epsilon}^2\sum_{i=1}^N (\hat{\nabla}c_i \otimes \hat{\nabla} c_i) \\
\label{eq:scaledD}\hat{\mathbb{D}}(\hat{\mathbf{v}}) &=& (\hat{\nabla} \otimes \hat{\mathbf{v}})+(\hat{\nabla}\otimes\hat{\mathbf{v}})^T \enskip ,
\end{eqnarray} 
respectively, whereas the dimensionless amplitudes are
\begin{equation}
\hat{a} = \frac{3\sigma_0\delta_0}{D_0^2\rho} \quad \text{and} \quad \hat{\eta} = \frac{\eta}{D_0\rho} \enskip .
\end{equation}
Finally, the incompressibility condition simply becomes
\begin{equation}
\label{eq:dyns3}
\hat{\nabla}\cdot\hat{\mathbf{v}} = 0 \enskip .
\end{equation}

\section{Numerical method}

The system of dynamic equations described by (\ref{eq:dyns1}), (\ref{eq:dyns2}) and (\ref{eq:dyns3}) are solved numerically on a fully periodic 2-dimensional domain by using an operator-splitting based quasi-spectral semi-implicit time stepping scheme \cite{Tegze20091612} combined with the spectral Chorin's projection method as follows. The dynamic equations can be re-written in the form
\begin{equation}
\label{eq:solve}
\frac{\partial \mathbf{y}}{\partial t} = \mathbf{f}(\mathbf{y},\nabla\mathbf{y}) \enskip ,
\end{equation} 
where $\mathbf{y}=(c_1,c_2,\dots,c_n,v_x,v_y)$, and $f\mathbf{f}(\mathbf{y},\nabla\mathbf{y})$ is the (generally nonlinear) right-hand side. $f(\mathbf{y},\nabla\mathbf{y})$ is calculated at time point $t$, while $\partial y_i/\partial t$ is discretized simply as
\begin{equation}
\label{eq:disct}
\frac{\partial y_i}{\partial t} \approx \frac{y_i^{t+\Delta t}-y_i^t}{\Delta t} \enskip .
\end{equation}
Next, we add the general linear term $\hat{s}[y_i]= \sum_{i=1}^\infty (-1)^i s_i \nabla^{2i} y_i$ (where $s_i \geq 0$) to both sides of Eq. (\ref{eq:solve}). We consider this term at $t+\Delta t$ at the left-hand side, but at $t$ on the right-hand side of the equation. This concept, together with Eq. (\ref{eq:disct}) results in the following, explicit spectral time stepping scheme:
\begin{equation}
\label{eq:timestep}
y_i^{t+\Delta t}(\mathbf{k}) = y_i^t(\mathbf{k}) + \frac{\Delta t}{1+s_i(\mathbf{k})\Delta t} \mathcal{F}\{f_i[\mathbf{y}^t(\mathbf{r}),\nabla\mathbf{y}^t(\mathbf{r})]\} \enskip ,
\end{equation}
where $s_i(\mathbf{k})=\sum_{j=1}^\infty s^{(i)}_j (\mathbf{k}^2)^j$, and $\mathcal{F}\{.\}$ stands for the Fourier transform. The \textit{splitting constants} $\{s^{(i)}_j\}$ must be chosen so that Eq. (\ref{eq:timestep}) is stable. Suitable splitting constants can be found by expanding the right-hand side of the differential equations, then identifying terms of the form $(-1)^{n+1} f(\mathbf{y})\nabla^{2\,n}y_i$ in the equation for $y_i$. $\max\{0,\max\{f(\mathbf{y})\}\}$ then provides a theoretical splitting constant $\tilde{s}_n^{(i)}$. Since the equations are coupled and highly non-linear, a unique experimental multiplier $s$ is applied, i.e. the splitting constants are chosen as $s_n^{(i)}:=s\,\tilde{s}_n^{(i)}$. In our case, we used $s=5$. 

Considering the Navier-Stokes equation, note that the new velocity field $\mathbf{v}^{t+\Delta t}(\mathbf{r})$ does not satisfy Eq. (\ref{eq:dyns3}) in general. Introducing $\mathbf{v}^{t+\Delta t}:=\mathbf{v}^*+\delta \mathbf{v}$, where $\mathbf{v}^*$ is calculated from Eq. (\ref{eq:timestep}), and the correction is given in the form $\delta \mathbf{v}:=\nabla s(\mathbf{r})$, where $s(\mathbf{r})$ is a scalar field, and using Eq. (\ref{eq:dyns3}) yields the spectral solution
\begin{equation}
\label{eq:scorrv}
\delta \mathbf{v} (\mathbf{k}) = - \frac{\mathbf{k} \otimes \mathbf{k}}{\mathbf{k}^2} \mathbf{v}^*(\mathbf{k}) \enskip .
\end{equation}
Using Eqns. (\ref{eq:timestep}) and (\ref{eq:scorrv}), the velocity field is then generated by the following sequence:
\begin{eqnarray}
\label{eq:velocity1}\mathbf{v}^*(\mathbf{k}) &=& \mathbf{v}^t(\mathbf{k}) + \frac{\Delta t}{1+s_v(\mathbf{k})\Delta t } \mathcal{F}\{\mathbf{f}^t(\mathbf{r})\} \\
\label{eq:velocity2}\mathbf{v}^{t+\Delta t}(\mathbf{k})&=& \left[ \mathbb{I} - \mathbb{P}(\mathbf{k}) \right] \cdot \mathbf{v}^*(\mathbf{k}) \enskip ,
\end{eqnarray}
where $s_v(\mathbf{k})$ is a splitting function emerging from the viscous stress, $\mathbf{f}^t(\mathbf{r})=\nabla \cdot \hat{\mathbb{P}}$, where $\hat{\mathbb{P}}$ defined by Eq. (\ref{eq:scaledP}), while $\mathbb{P}(\mathbf{k})=(\mathbf{k}\otimes\mathbf{k})/\mathbf{k}^2$  is the operator generating the divergent part of a vector field. Indeed, in Eq. (\ref{eq:velocity2}) $\mathbb{I}-\mathbb{P}(\mathbf{k})$ eliminates the divergence of $\mathbf{v}^*$.

It is important to note that our numerical scheme is \textit{unbounded}, meaning that the spatial solution $c_i(\mathbf{r},t)$ might become negative or greater than 1 because of numerical errors. Nevertheless, the construction of the free energy functional and the modified Bollada-Jimack-Mullis mobility matrix ensure that no artificial modification of the solution is needed after a time step, as discussed before. Instead, the system naturally finds the bulk states and the two-component interfaces. Finally we mention, that the generalized Chorin's projection method presented here is compatible with equilibrium. In equilibrium the diffusion fluxes vanish, i.e. $\mathbf{J}_i=0$ for $i=1\dots N$, resulting in $\dot{\mathbf{c}}=0$. Furthermore, $\nabla\cdot\mathbb{A}$ becomes the gradient of a scalar function in equilibrium, which is then eliminated by the Chorin's projection method (i.e. no flow is generated). Since the viscous term vanish for a homogeneous velocity field, $\mathbf{v}(\mathbf{r})=const$ is the general equilibrium solution.

\begin{figure}[t]
\begin{center}
\includegraphics[width=0.49\linewidth]{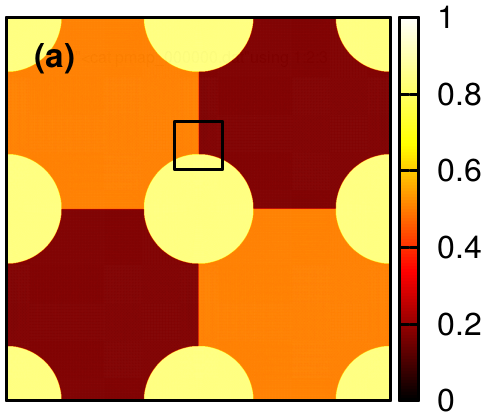}
\includegraphics[width=0.49\linewidth]{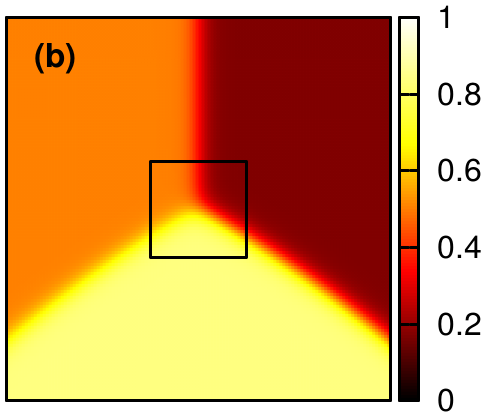}\\
\includegraphics[width=0.49\linewidth]{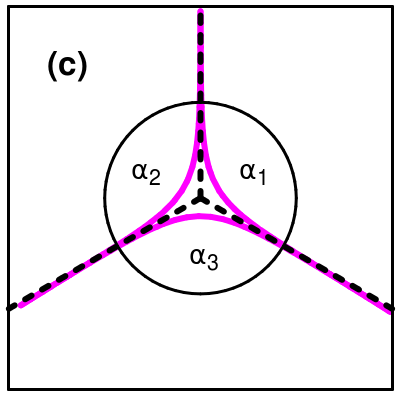}
\includegraphics[width=0.49\linewidth]{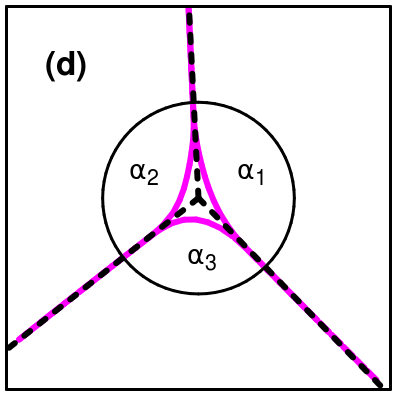}\\
\caption{Contact angle measurement in a ternary system: (a) Initial condition, and (b) converged (equilibrium) solution in a symmetric system in the area indicated by the black square on panel a. On both panels $\sum_{i=1}^3 c_i(\mathbf{r})[(i-0.5)/3]$ is shown. (c) Contour lines [$c_i(\mathbf{r})=0.5$ for $i=1\dots 3$] of the fields at a trijunction in the area indicated by the black square on panel b, and (d) the same as panel c in case of asymmetric system.}
\end{center}
\end{figure}

\section{Results}

The numerical simulations were performed on a 2-dimensional, uniform rectangular grid with spatial resolution $h=0.5$ and different time steps. The physical parameters were chosen to model realistic binary, ternary and quaternary (4-component) systems mimicking the oil/water/CO$_2$ interfaces. The scales then read $\rho=1000$ kg/m$^3$, $D_0=5\times10^{-10}$ m$^2$/s, $\sigma_0=50$ mJ/m$^2$, $\delta_0=1.25$ \AA, and 
\begin{equation}
\label{eq:scaledvisc}
\eta(\mathbf{c}):=\eta_0 \sum_{i=1}^N c_i x_i \enskip ,
\end{equation}
where $x_i=\eta_i/\eta_0$, and the viscosity scale reads $\eta_0=1$ mPas.

\begin{figure}[t]
\begin{center}
\includegraphics[width=0.49\linewidth]{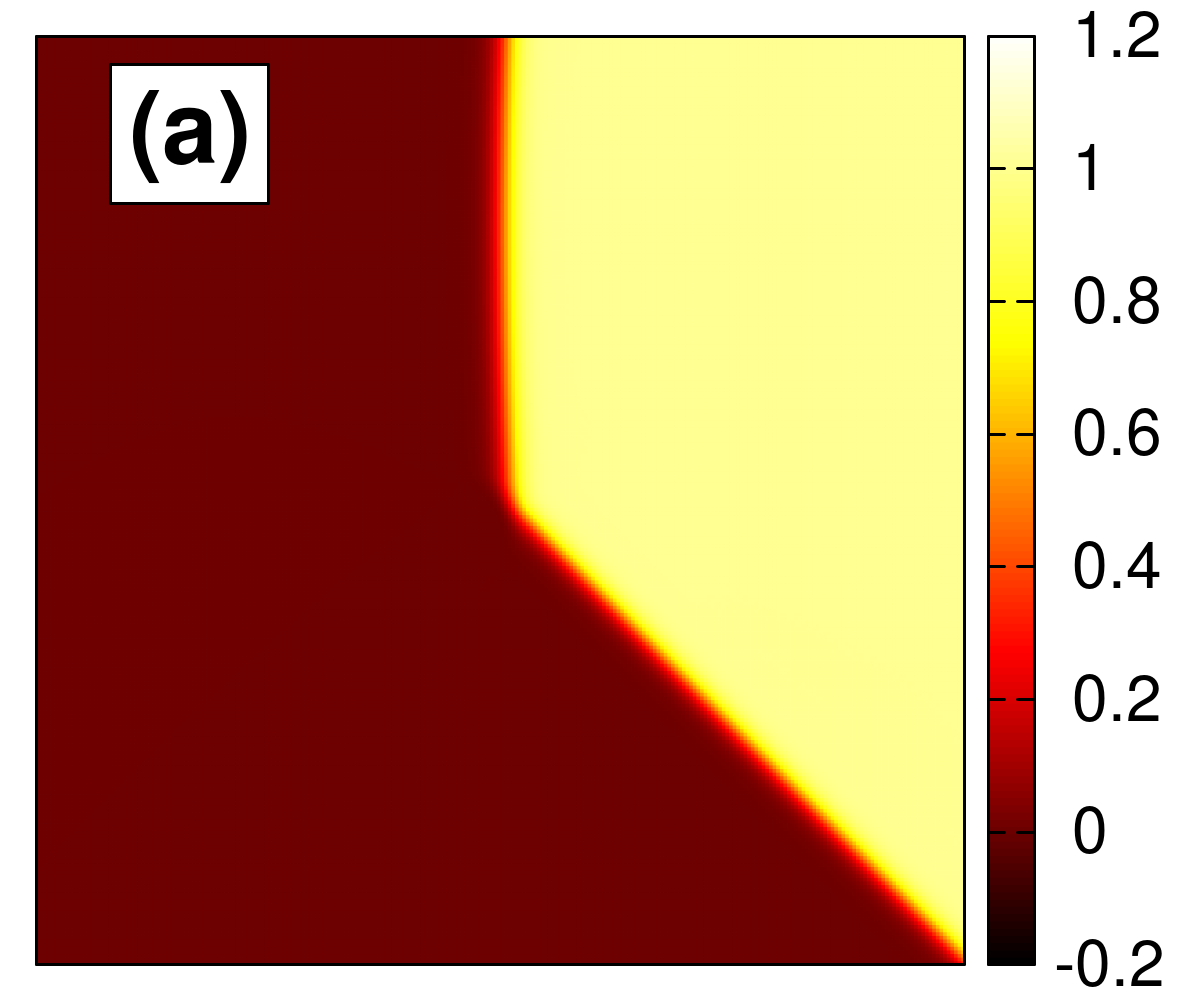}
\includegraphics[width=0.49\linewidth]{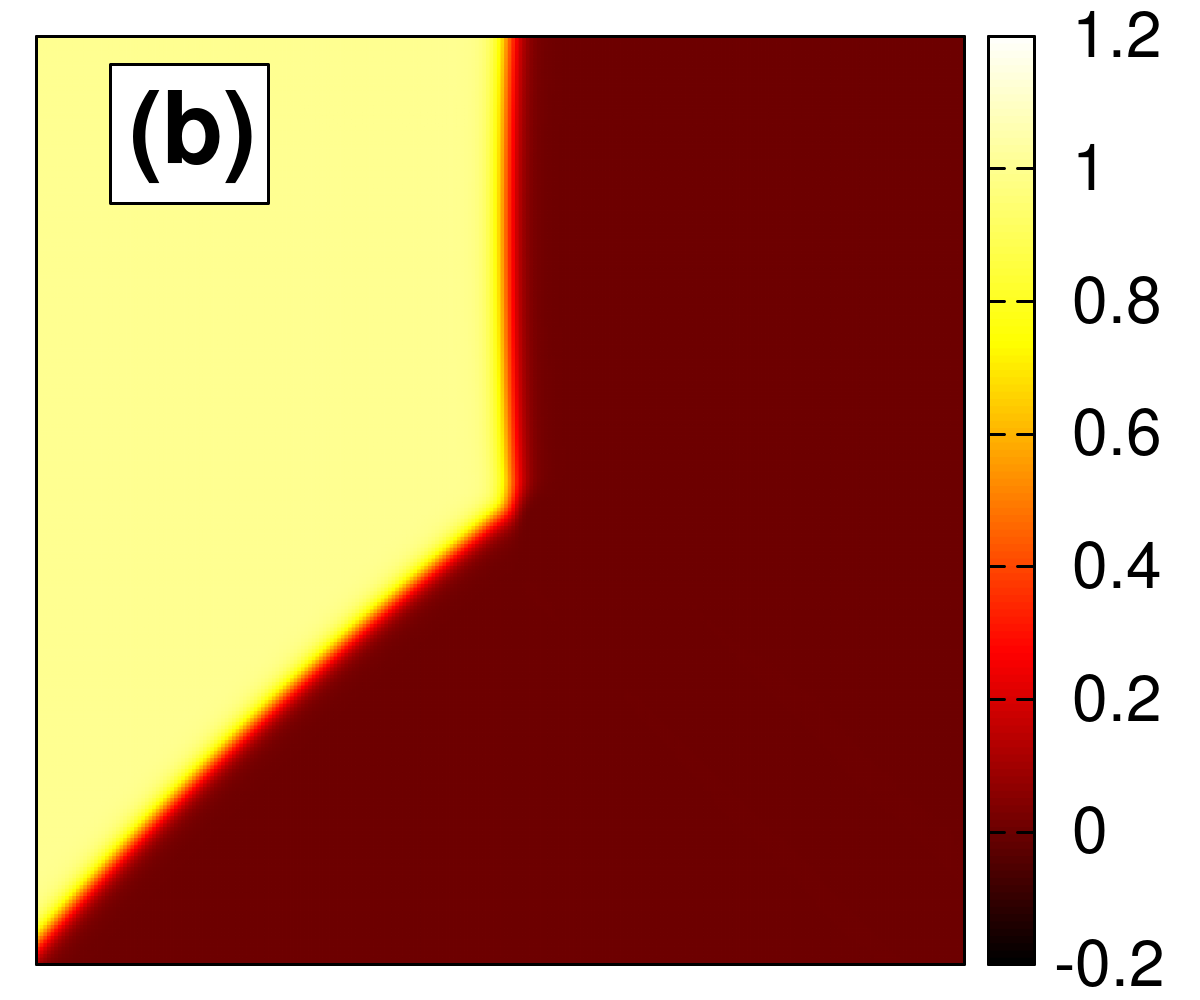}\\
\includegraphics[width=0.49\linewidth]{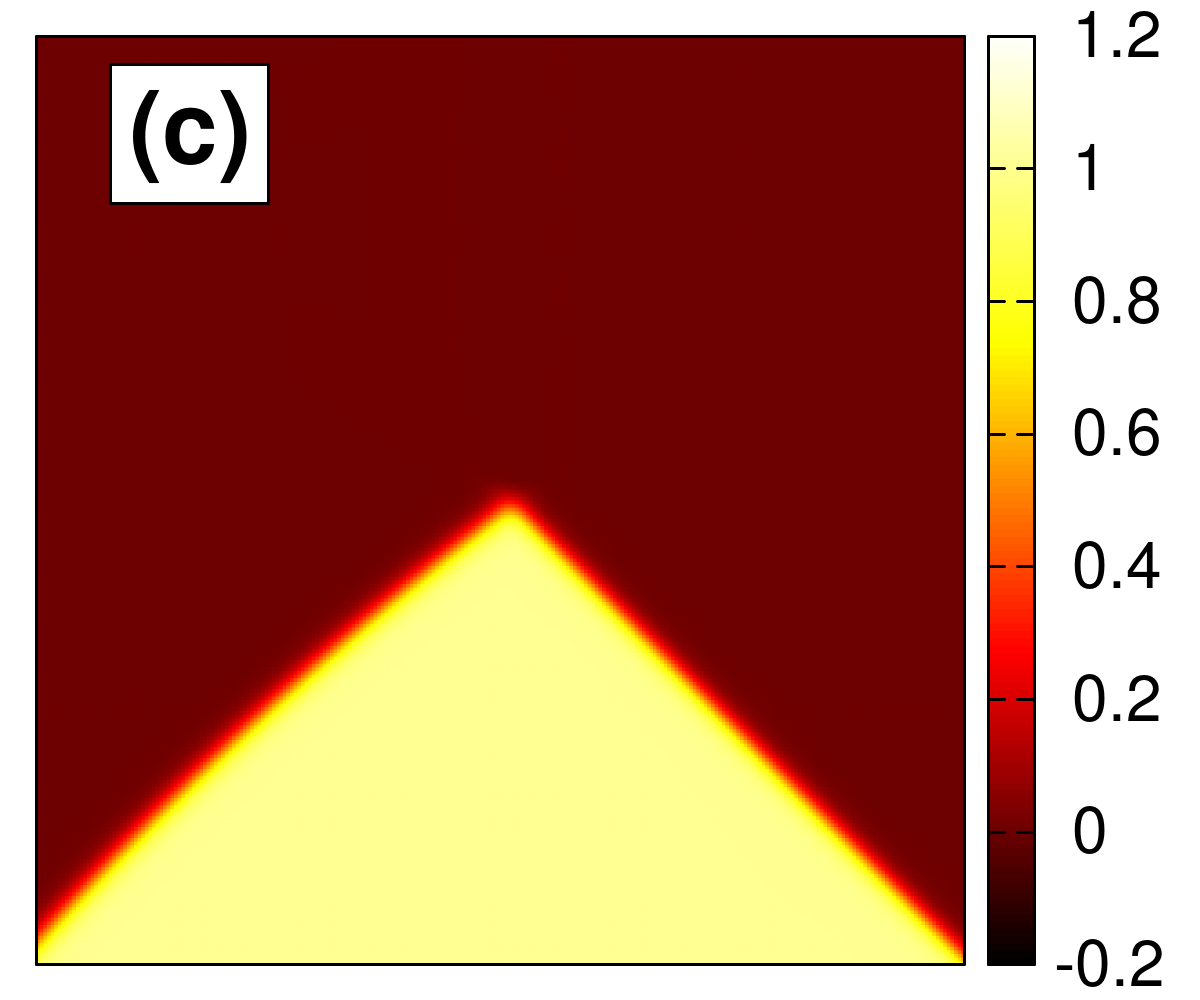}
\includegraphics[width=0.49\linewidth]{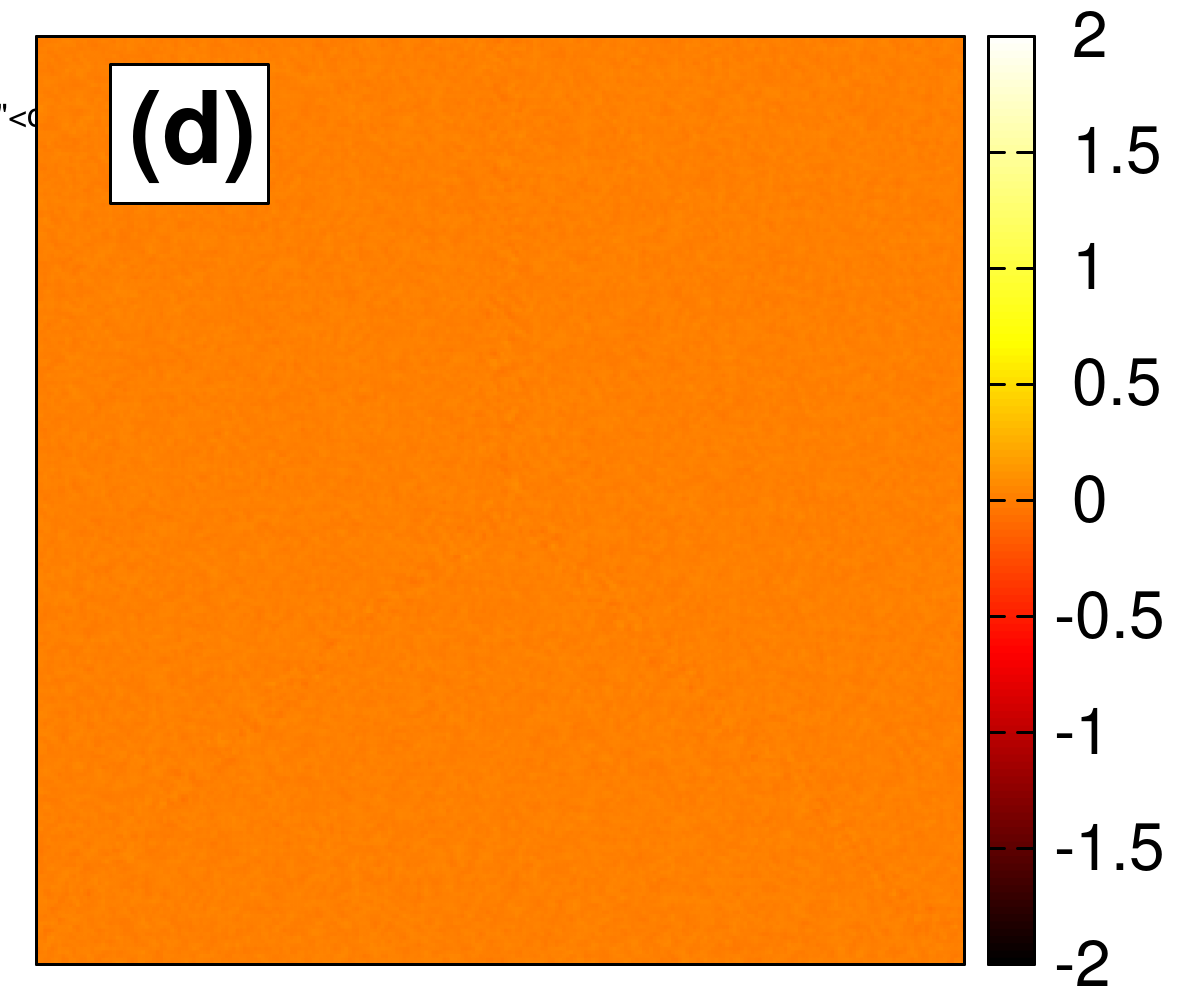}\\
\caption{Spatial distribution of the individual components (a-c) in the vicinity of the equilibrium trijunction in an asymmetric ternary system, and (d) error of the local sum of the variables, $e:=10^{14}\left[\sum_{i=1}^3 c_i(\mathbf{r})-1\right]$. Note that the third component is not present at the binary interfaces, while the error of the local sum is negligible.}
\end{center}
\end{figure}

\subsection{Contact angles}

The validation of the model started with equilibrium contact angle measurements in both symmetric ($\hat{\sigma}_{ij}=\hat{\delta}_{ij}=1$) and asymmetric systems. As discussed in Section III, the function $h(c_i,c_j)=|c_i/(1-c_i)||c_j/(1-c_j)|$ in Eq. (\ref{eq:scaledJ}) might generate "dangerous" solutions (i.e. stationary solutions which do not represent equilibrium), therefore, the dynamic equations were solved by applying $h(c_i,c_j) \equiv 1$ (and $\hat{\kappa}_{ij}=1/2$) in this case. Since we are interested exclusively in equilibrium, but not the time evolution of the system, this step does not influence the results. The initial condition for the velocity field was $\mathbf{v}(\mathbf{r},0)=0$, while the initial distribution of the components is shown in Fig. 4(a). For better visualization, $h_3(\mathbf{r},t):=\sum_{i=1}^3 c_i(\mathbf{r},t)[(i-0.5)/3]$ is shown, thus indicating bulk components at $h=1/6,1/2$ and $5/6$ for $i=1,2$ and $3$, respectively. The calculations were performed on a $1024\times1024$ grid with time step $\Delta t=0.001$. After $10^6$ time steps the flow field vanished, and the system practically reached equilibrium [the convergence criterion for equilibrium was $\bar{v}:=1/(N_x N_y)\sum_{i,j}\sqrt{\mathbf{v}_{i,j}^2} < 10^{-4}$ for the average velocity, which corresponds to 1 pixel shift in the solution in $10^{6}$ time steps]. The amplitude of the triplet term was $A_3=0$ and $1/2$  in the symmetric and asymmetric system, respectively. 

In order to measure the contact angles at a trijunction, we plotted the $c_i(\mathbf{r})=1/2$ contours for $i=1,2$ and $3$, as shown in Fig. 4(c), then fitted straight lines (dashed in the figure) for the unperturbed binary interfaces ("far" from the trijunction). The crossing point of these lines defines the trijunction point. As expected, the contact angle $\alpha_1=\alpha_2=\alpha_3=120^o$ was detected in the symmetric system. In contrast, asymmetric systems establish different contact angles. For instance, for the interface tensions $\hat{\sigma}_{12}=1.2$, $\hat{\sigma}_{13}=1.0$ and $\hat{\sigma}_{23}=0.8$ (the corresponding interface thicknesses were $\hat{\delta}_{12}=1.1$, $\hat{\delta}_{13}=0.9$ and $\hat{\delta}_{23}=1.0$, respectively), the theoretical contact angles can be determined from the condition of mechanical equilibrium, yielding:
\begin{eqnarray}
\alpha_1^0 &=& \pi - \cos^{-1} \left( \frac{\hat{\sigma}_{12}^2+\hat{\sigma}_{13}^2-\hat{\sigma}_{23}^2}{2\hat{\sigma}_{12}\hat{\sigma}_{13}} \right) = 138.6^o \quad \\
\alpha_2^0 &=& \pi - \cos^{-1} \left( \frac{\hat{\sigma}_{12}^2+\hat{\sigma}_{23}^2-\hat{\sigma}_{13}^2}{2\hat{\sigma}_{12}\hat{\sigma}_{23}} \right) = 124.23^o \quad \\
\alpha_3^0 &=& \pi - \cos^{-1} \left( \frac{\hat{\sigma}_{13}^2+\hat{\sigma}_{23}^2-\hat{\sigma}_{12}^2}{2\hat{\sigma}_{13}\hat{\sigma}_{23}} \right) = 97.181^o \enskip . \quad
\end{eqnarray}
From the simulation, the contact angles $\alpha_1=137.3^o$, $\alpha_2=126.37^o$, and $\alpha_3=96.33^o$ have been measured [see Fig 4(d)], showing then $1.7\%$ maximal relative error compared to the theoretical values, which can be attributed to the uncertainty of the measurement.

Figure 5 shows the individual compositions (panels a-c) and the sum of the fields (panel d) in the neighborhood of the trijunction displayed in Figure 4(d). The spatial distribution of the individual fields demonstrate the effect of the triplet term. In accordance with Figure 3(b) and 4(d), all of the two-component interfaces are practically free of the third component. Furthermore, Figure 5(d) shows that the error of the local sum of the variables is in the range of the truncation error of double precision floating point numbers.

The calculations were repeated in an asymmetric 4-component (quaternary) system as well (see Fig 6), with $\hat{\sigma}_{12}=1.0$, $\hat{\sigma}_{13}=1.1$, $\hat{\sigma}_{14}=0.75$, $\hat{\sigma}_{23}=0.9$, $\hat{\sigma}_{24}=1.25$ and $\hat{\sigma}_{34}=1.0$. The interface thicknesses were equal, i.e. $\hat{\delta}_{ij}=1.0$ was used, while the amplitude of the triplet term was $A_3=1$. The contact angle measurements resulted in less than $1.5\%$ relative error again compared to the theoretical values for all the 4 different trijunctions [illustrated in Fig 6(c)-(f)]. According to our experience, the unperturbed binary planar interfaces contain no additional components, similarly to the ternary case.

\begin{figure}[t]
\begin{center}
\includegraphics[width=0.49\linewidth]{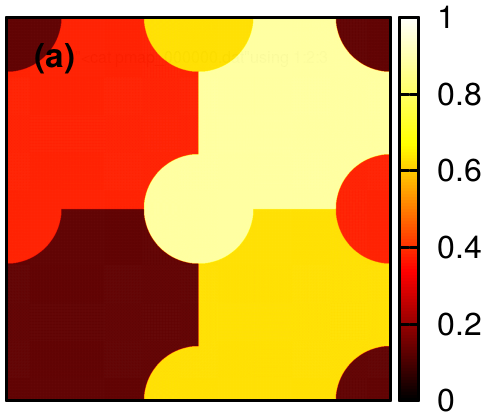}
\includegraphics[width=0.49\linewidth]{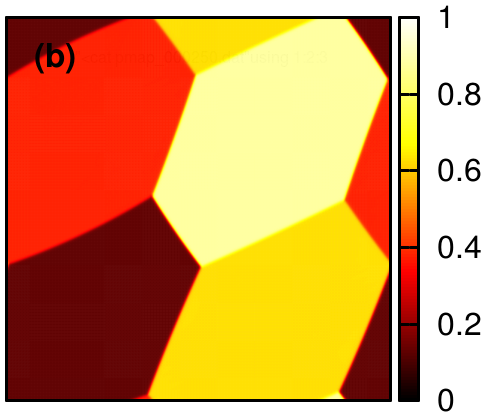}\\
\includegraphics[width=0.49\linewidth]{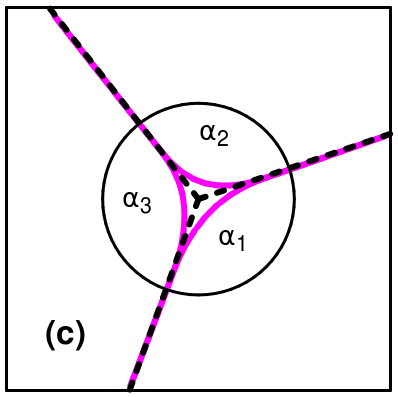}
\includegraphics[width=0.49\linewidth]{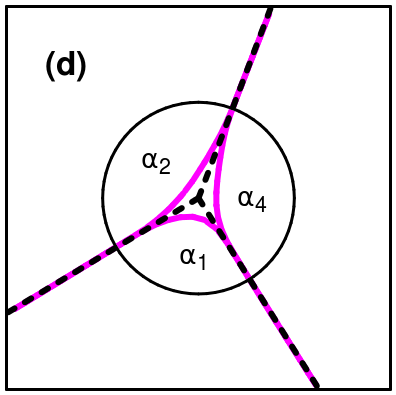}\\
\includegraphics[width=0.49\linewidth]{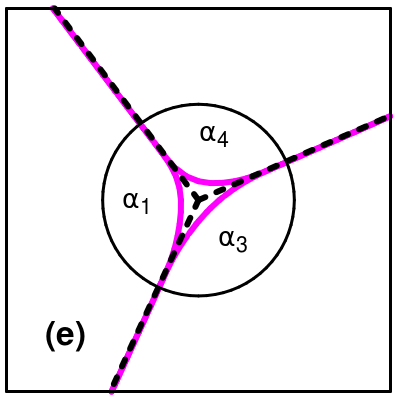}
\includegraphics[width=0.49\linewidth]{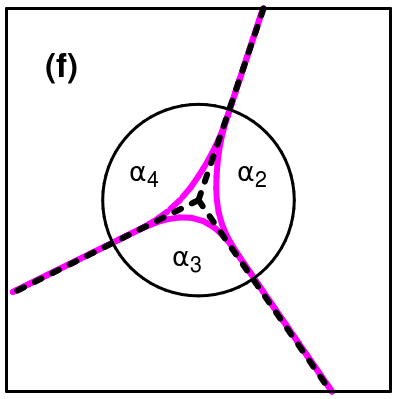}\\
\caption{Contact angles in an asymmetric quaternary system (for parameters, see the main text): (a) initial condition, (b) equilibrium state, (c)-(d) contour lines for the fields in the vicinity of the $4$ different trijunctions from panel b, analogously to Fig 4. On panels a and b, $h_4(\mathbf{r},t)=\sum_{i=1}^4 [ (i-1/2)/4]c_i(\mathbf{r},t)$ is shown.}
\end{center}
\end{figure}

\subsection{Spinodal decomposition}

Since we're now interested in the time evolution of the system, the modified Bollada-Jimack-Mullis matrix defined by Eq. (\ref{eq:BJMmobmat}) is used henceforth.

\subsubsection{Binary system}

\begin{figure}[t]
\begin{center}
\includegraphics[width=0.49\linewidth]{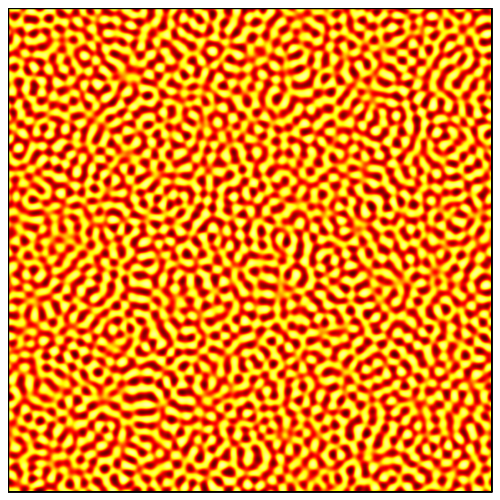}
\includegraphics[width=0.49\linewidth]{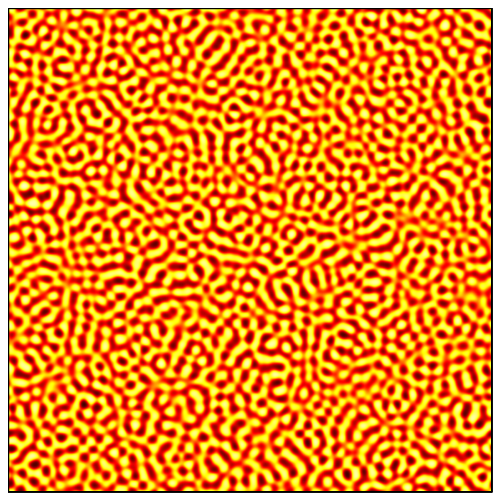}\\
\includegraphics[width=0.49\linewidth]{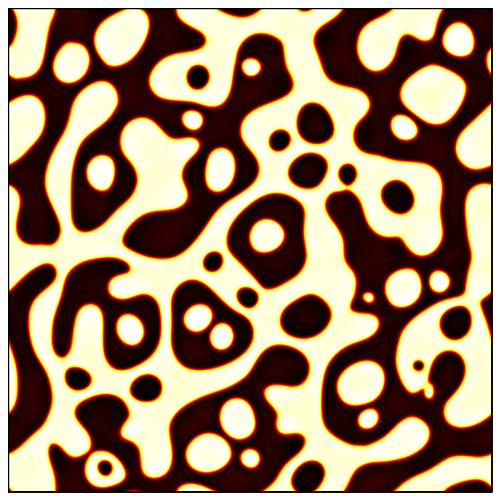}
\includegraphics[width=0.49\linewidth]{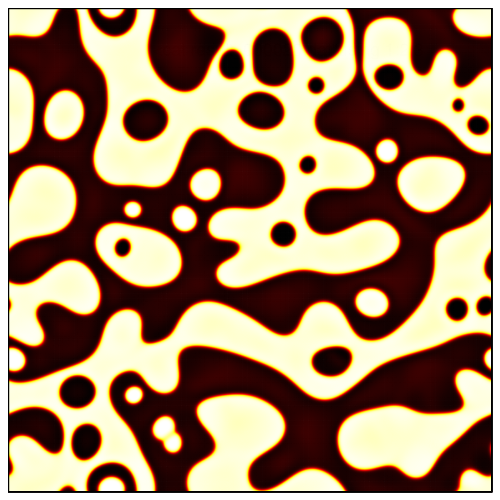}\\
\includegraphics[width=0.49\linewidth]{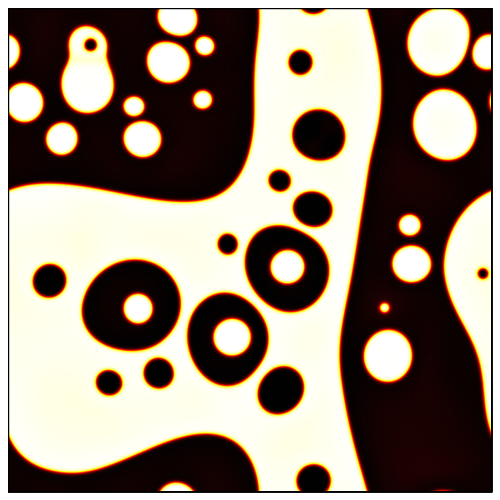}
\includegraphics[width=0.49\linewidth]{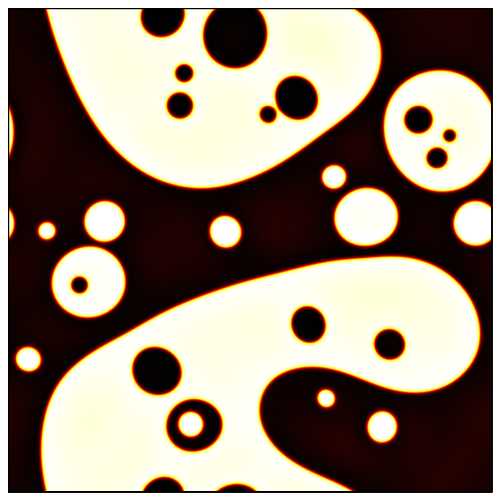}\\
\caption{Pattern coarsening during liquid-flow assisted spinodal decomposition of a binary liquid, as predicted by the Ginzburg-Landau theory of surfactant assisted liquid phase separation of T\'oth and Kvamme (left column), and the present model (right column). The snapshots of the simulations were taken at $t=62.5,125,$ and $250$, respectively (from top to bottom).}
\end{center}
\end{figure}

Spinodal decomposition was studied first in the binary limit. Technically we performed calculations in a ternary system, where the third component was set to 0 initially, i.e. $c_3(\mathbf{r},0)=0$ was used. In this case, the dynamic equations, together with the Navier-Stokes equation naturally reduce to the dynamic equations of a traditional, one order parameter, flow assisted Cahn-Hilliard system. Therefore, the reference calculation was based on the surfactant assisted liquid phase separation model of T\'oth and Kvamme for incompressible liquid flow, in the surfactant-free case. The dynamic equations read
\begin{eqnarray}
\dot{\phi} &=& \nabla^2 [(\phi^3-\phi)-2\nabla^2 \phi] \\
\dot{\mathbf{v}} &=& \nabla \cdot (\mathbb{A}+\mathbb{D}) \\
\mathbb{A} &=& - 2\, \tilde{w}\,( \nabla \phi \otimes \nabla \phi) \\
\mathbb{D} &=& \tilde{\mu} [(\nabla \otimes \mathbf{v})+(\nabla\otimes\mathbf{v})^T] \\
0 &=& \nabla \cdot \mathbf{v} \enskip . 
\end{eqnarray}
The transformation of the fields read $c_1=(1+\phi)/2$ and $c_2=(1-\phi)/2$, yielding $\hat{\kappa}_{12}^0=1$, $\hat{\eta}=\hat{\eta}_0(c_1\,x_1+c_2\,x_2)$ corresponding to $\tilde{\mu}=\tilde{\mu}_0[x_1(1+\phi)/2+x_2(1-\phi)/2]$ with $\hat{\eta}_0=\tilde{\mu}_0$, and $\hat{a}=4\tilde{w}$. We used $\tilde{\mu}_0=2857.0$, $x_1=1.0$ and $x_2=1633.0/\tilde{\mu}_0$ in Eq. (\ref{eq:scaledvisc}), and $\tilde{w}=1.73\times10^4$. The initial condition was $\phi(\mathbf{r},0)=A\,\mathcal{R}[-1,+1]$ [and $c_1(\mathbf{r},0)=0.5+(A/2)\mathcal{R}[-1,+1]$, correspondingly], where $\mathcal{R}[-1,1]$ is a uniformly distributed random number on $[-1,1]$, and $|A| \ll 1$. Since the homogeneous state $\phi=0$ (and $c_1=0.5$) represents unstable equilibrium, the system undergoes phase separation for $A \neq 0$. Since the implementation of the equations in solving the different models are different, we do not expect exactly the \textit{same} result from the same initial condition. Nevertheless, we are interested only in the characteristic behavior of the system. Therefore, we used different random numbers (but the same amplitude $A$) in setting up the initial conditions for $\phi$ and $c_1$. In this case $\Delta t=0.0025$ was chosen. Snapshots of the simulations are presented in Fig 7. It is quite clear that the patterns roughen similarly as a function of time in both cases, indicating that the dynamic equations of the present model reduce naturally to the conventional binary model. In addition, no appearance of the third component was detected in our model during the simulation, due to the Bollada-Jimack-Mullis type mobility matrix.

\begin{figure}[t!]
\begin{center}
\includegraphics[width=0.49\linewidth]{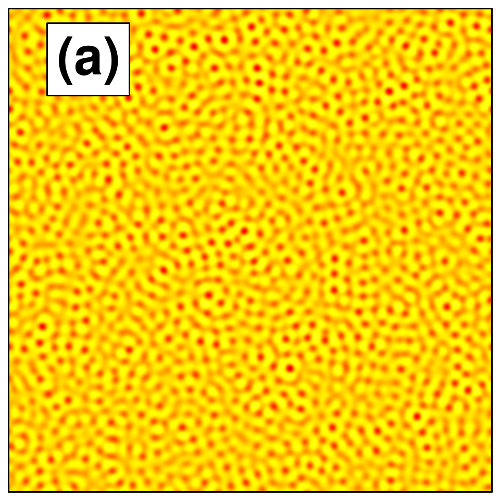}
\includegraphics[width=0.49\linewidth]{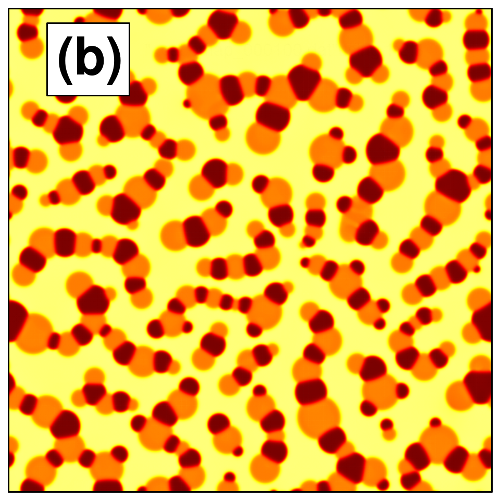}\\
\includegraphics[width=0.49\linewidth]{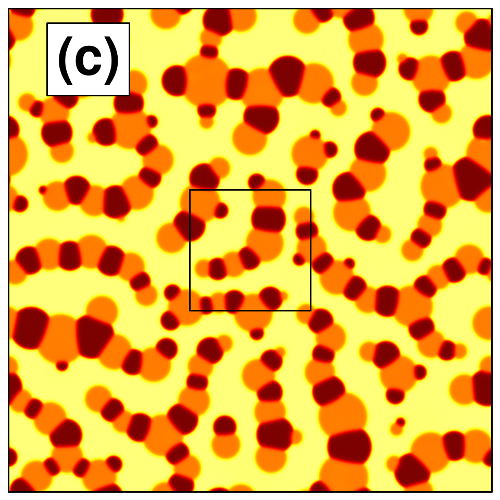}
\includegraphics[width=0.49\linewidth]{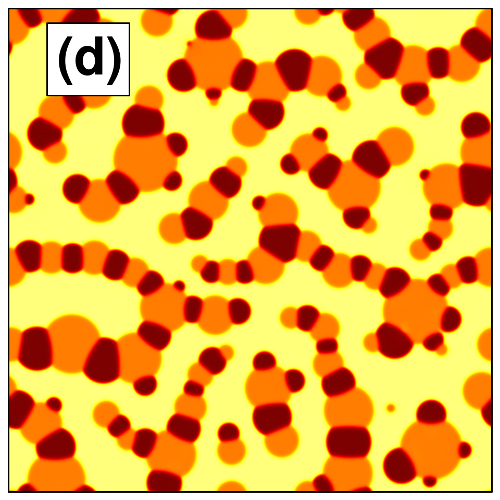}\\
\includegraphics[width=0.49\linewidth]{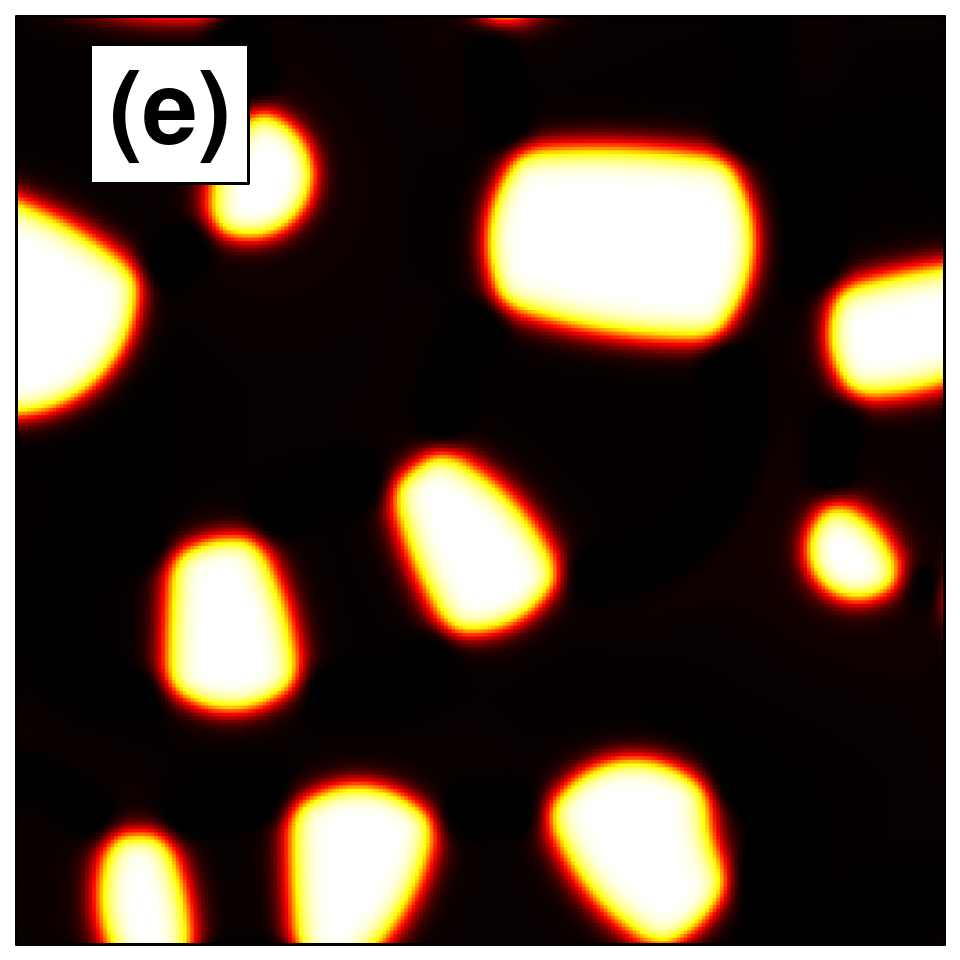}
\includegraphics[width=0.49\linewidth]{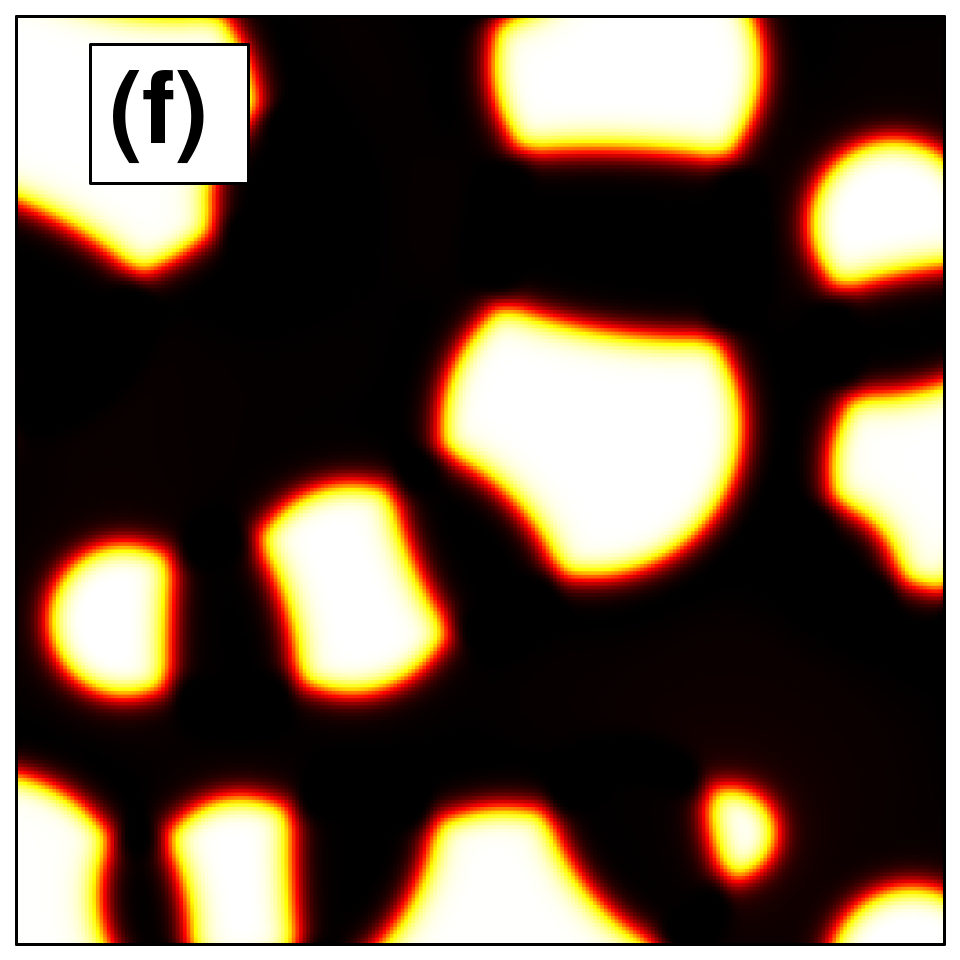}\\
\includegraphics[width=0.49\linewidth]{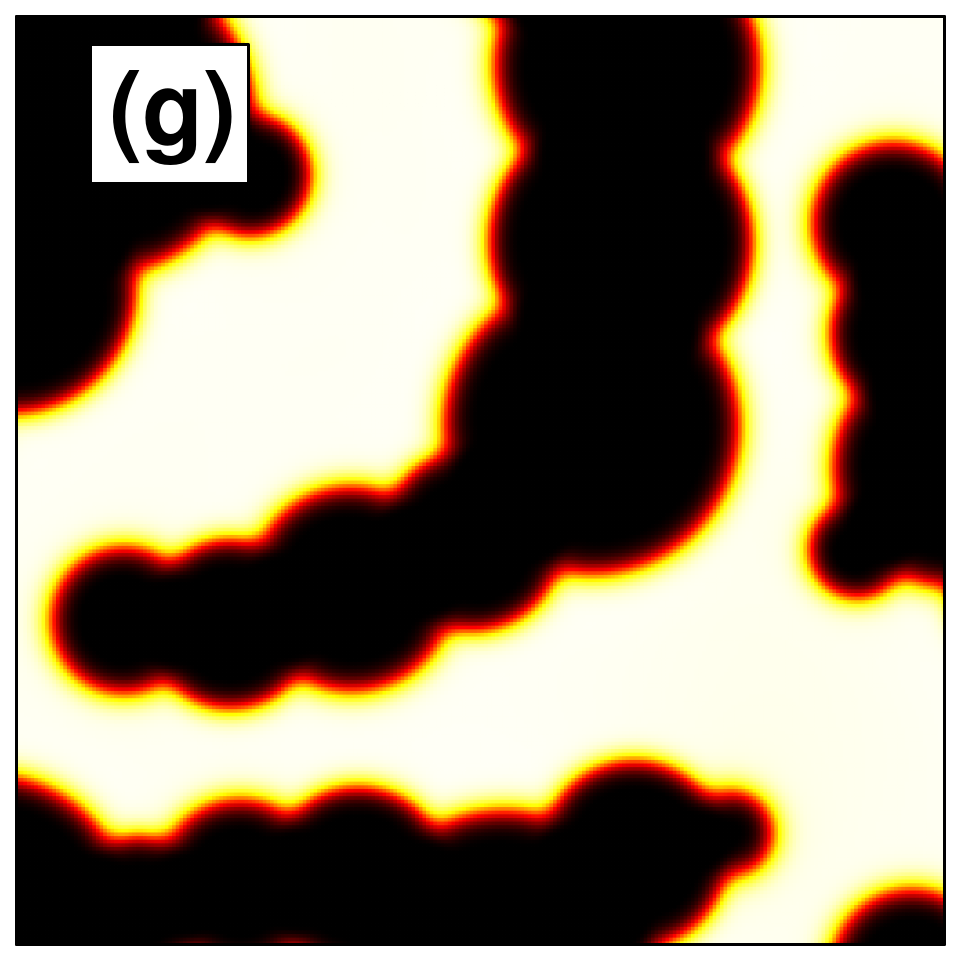}
\includegraphics[width=0.49\linewidth]{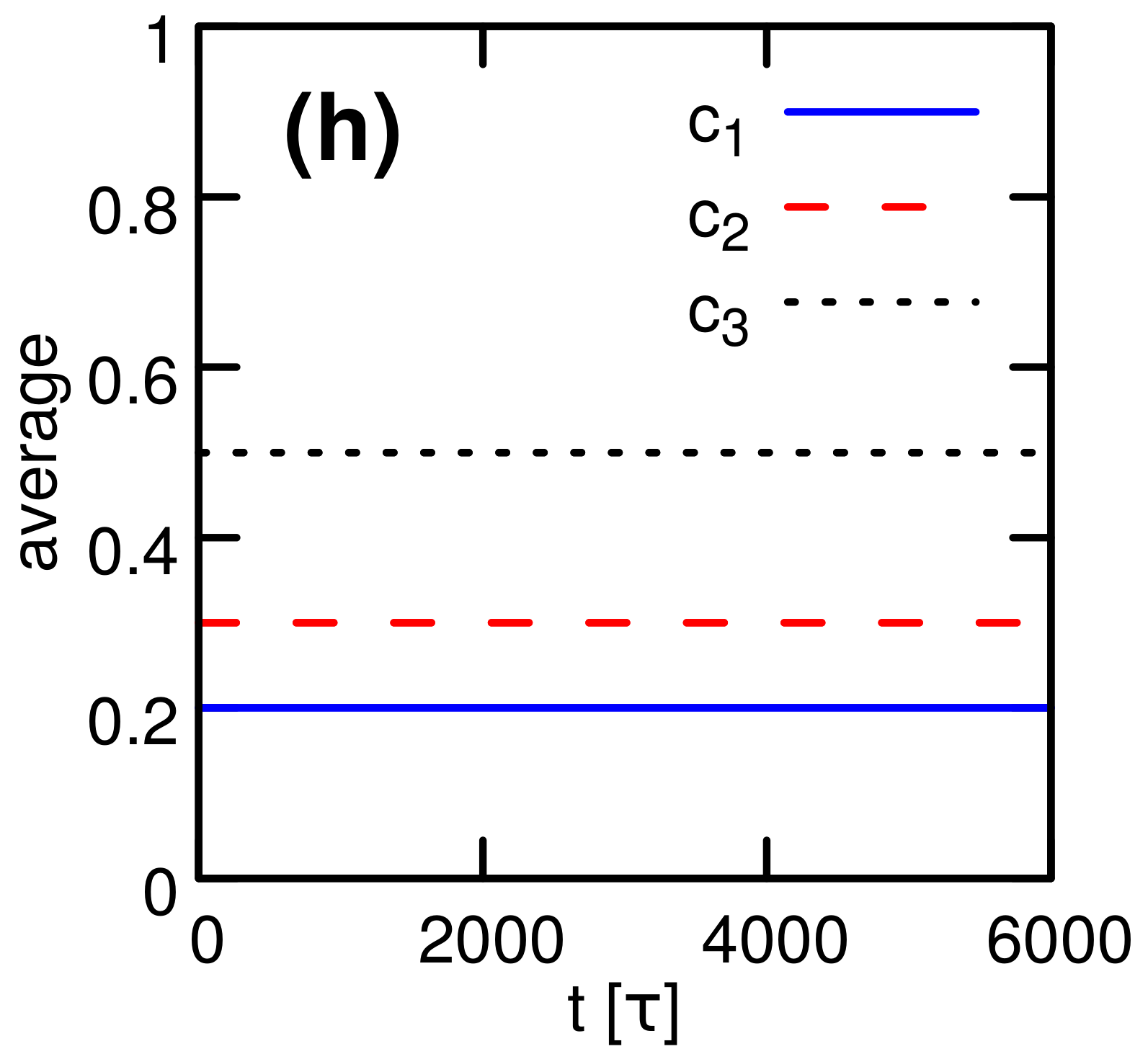}\\
\caption{Spinodal decomposition in an asymmetric ternary system. Snapshots of the simulation at $t=312.5,1250,3125$ and $6250$ (from panels a to d), respectively. Coloring is the same as in Figure 6. Panels e-g show the individual mass fractions $c_1(\mathbf{r},t)$, $c_2(\mathbf{r},t)$ and $c_3(\mathbf{r},t)$, respectively, in the area indicated by the black square on panel c. (Black corresponds to $c=0$ and white to $c=1$.) The time evolution of the total concentrations are shown by panel h, thus indicating global conservation for all components.}
\end{center}
\end{figure}

\begin{figure}
\begin{center}
\includegraphics[width=0.49\linewidth]{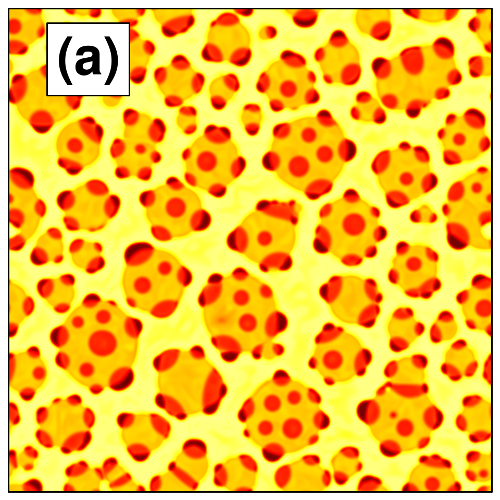}
\includegraphics[width=0.49\linewidth]{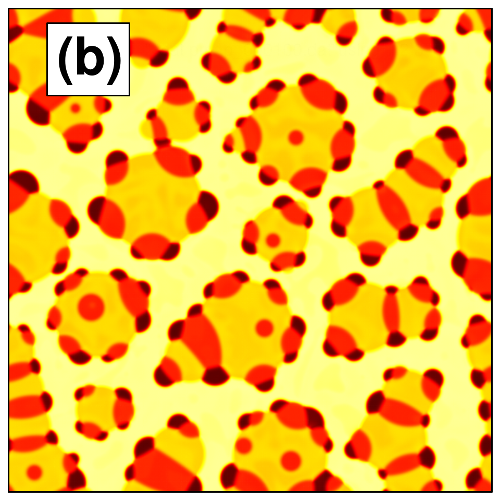}\\
\includegraphics[width=0.49\linewidth]{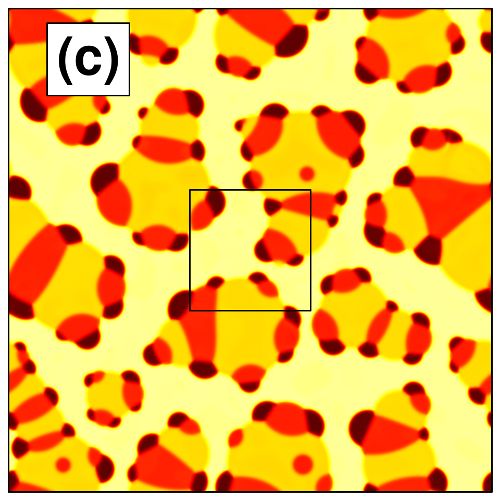}
\includegraphics[width=0.49\linewidth]{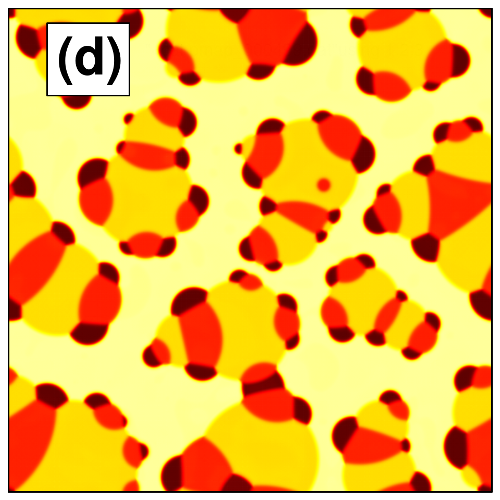}\\
\includegraphics[width=0.49\linewidth]{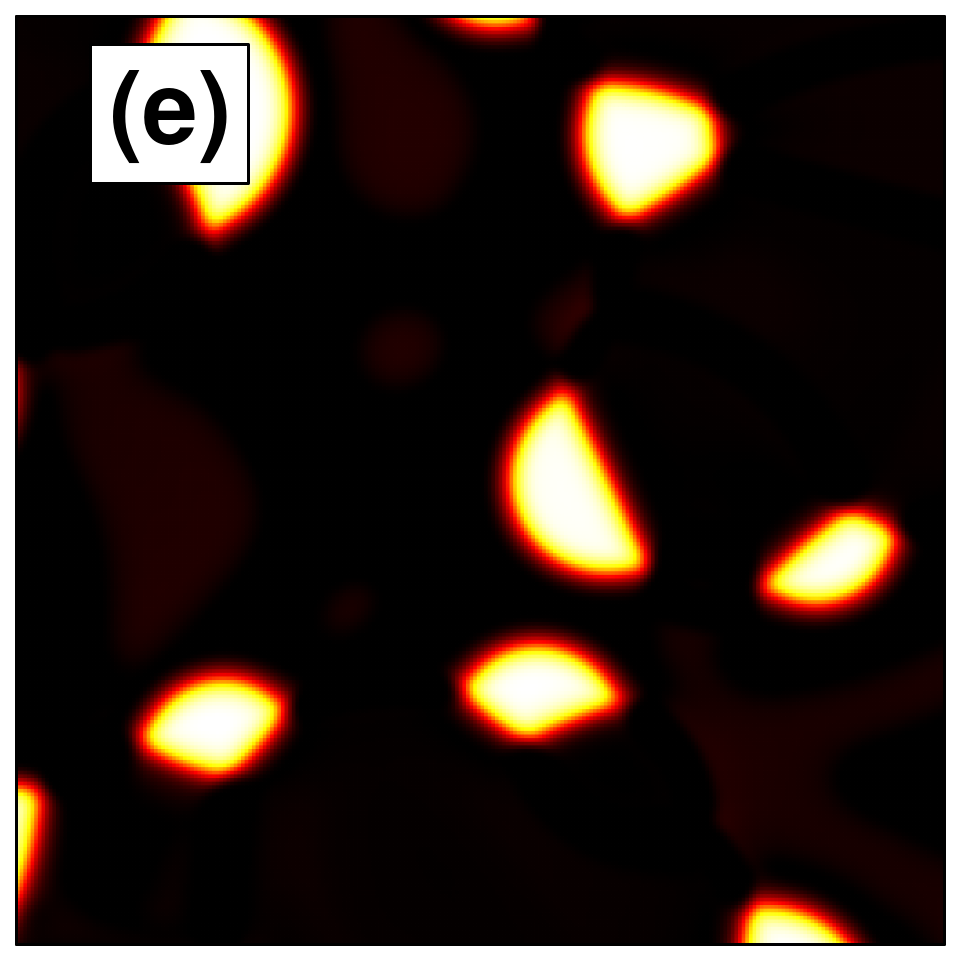}
\includegraphics[width=0.49\linewidth]{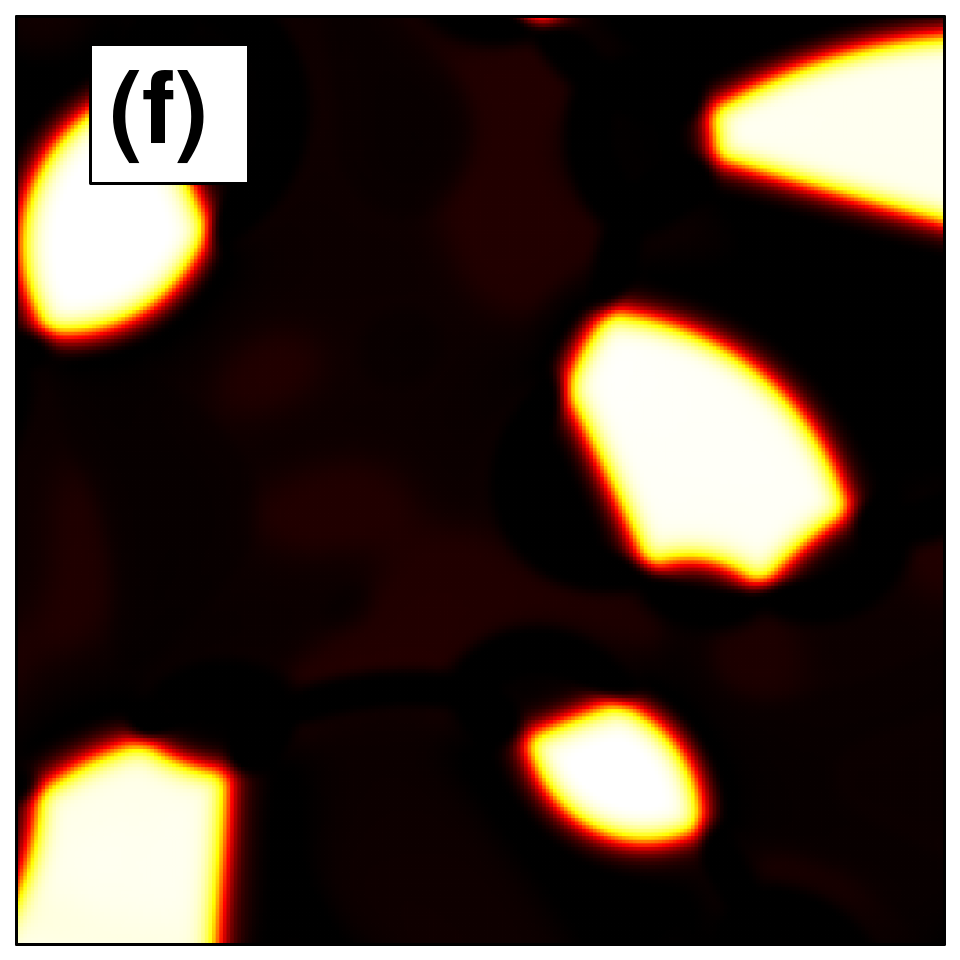}\\
\includegraphics[width=0.49\linewidth]{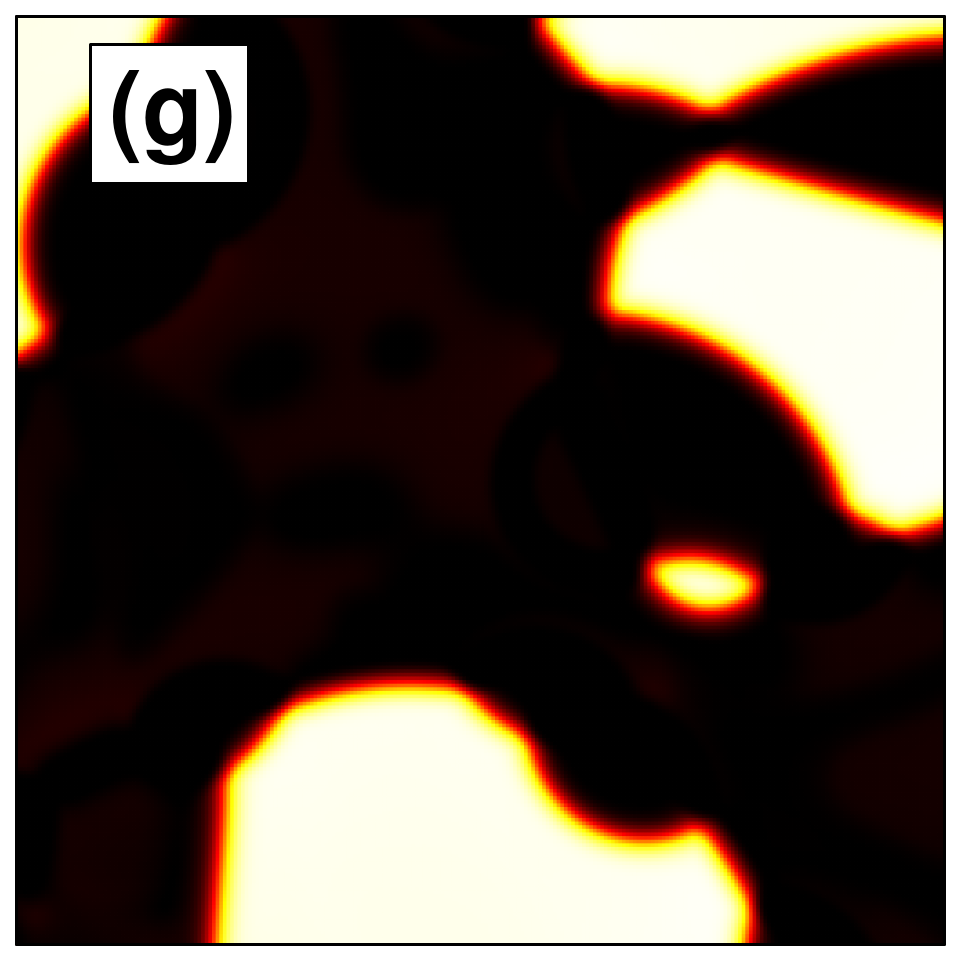}
\includegraphics[width=0.49\linewidth]{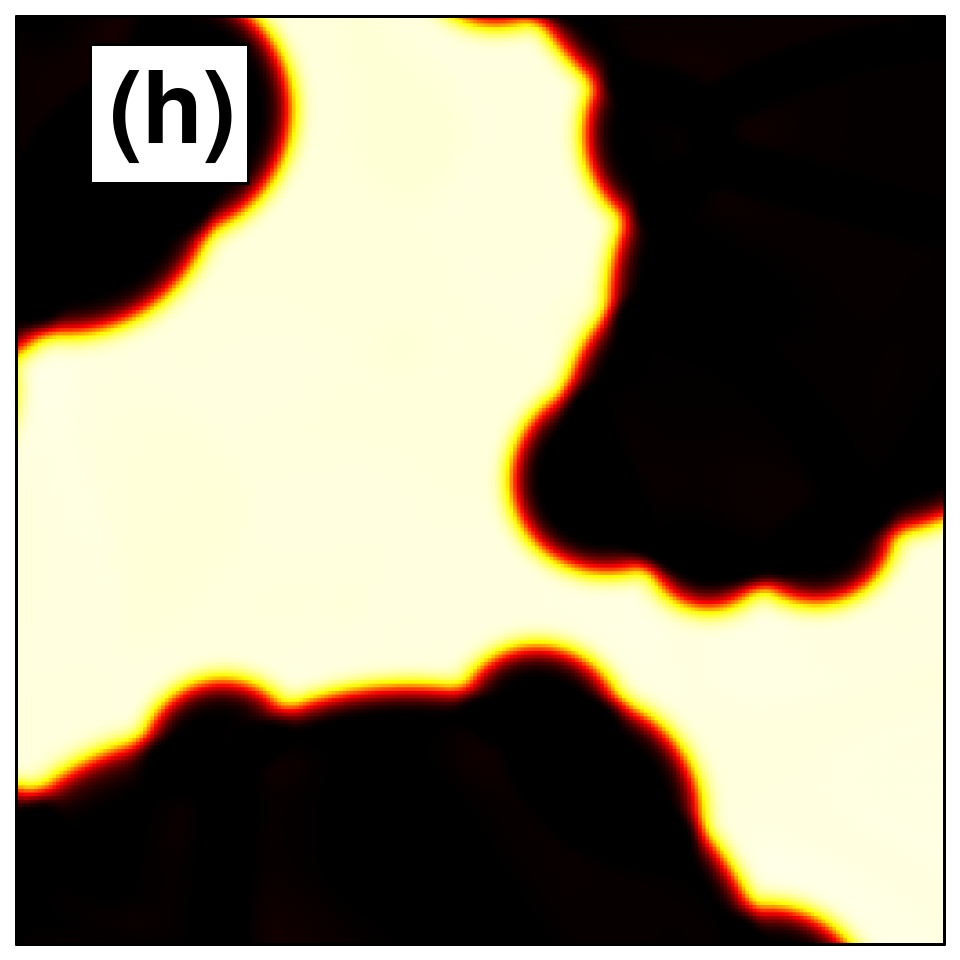}\\
\caption{Spinodal decomposition in an asymmetric quaternary (4-component) system. Snapshots at dimensionless times $t=312.5,1250,3125$ and $6250$, respectively. The individual fields $c_1(\mathbf{r},t)$, $c_2(\mathbf{r},t)$, $c_3(\mathbf{r},t)$ and $c_4(\mathbf{r},t)$ are shown in panels e-f in the black square indicated in panel c.}
\end{center}
\end{figure}

\subsubsection{Asymmetric ternary and quaternary flows}

In our first multi-component simulation an asymmetric ternary system was considered with dimensionless interfacial tensions $\hat{\sigma}_{12}=1.2$, $\hat{\sigma}_{13}=1.0$, and  $\hat{\sigma}_{23}=0.8$, and dimensionless interface thicknesses $\hat{\delta}_{12}=1.1$, $\hat{\delta}_{13}=0.9$ and $\hat{\delta}_{23}=1.0$. The amplitude of the triplet term was $A_3=1/2$, which was enough to stabilize the binary planar interfaces. The pairwise diffusion constants were also asymmetric, we used $\hat{D}_{12}=1.0$, $\hat{D}_{13}=2.0$, and $\hat{D}_{23}=0.5$, whereas the dimensionless viscosities in Eq. (\ref{eq:scaledvisc}) were $x_1=0.5$, $x_2=1.0$ and $x_3=2.0$, respectively. The initial condition reads $c_1(\mathbf{r},0)=0.2+A\,\mathcal{R}[-1,1]$, $c_2(\mathbf{r},0)=0.3+A\,\mathcal{R}[-1,1]$, and $c_3(\mathbf{r},0)=1-[c_1(\mathbf{r},0)+c_2(\mathbf{r},0)]$, where $A=0.01$ was chosen. The simulation has been performed on a $1024\times1024$ computational grid with $h=0.5$ and $\Delta t=0.005$. Snapshots of the simulation are shown in Fig. 8(a)-(d) at different dimensionless times. As one can see, the system is unstable in its initial state, and undergoes spinodal decomposition. Although the system is still far from equilibrium at $t=3125$, the individual fields of the components [see panels (e)-(g)] suggest, that the third component vanishes at the evolving binary interfaces. It is nevertheless important to mention, that pure binary interfaces exist only in equilibrium, while non-equilibrium curved interfaces may contain the third component. This effect is not prevented by applying a mobility matrix of the Bollada-Jimack-Mullis type, which is responsible only for preventing the appearance of a component when it is not present in a calculation \textit{at all} \cite{PhysRevB.92.184105}. Despite these, the third component tends to vanish at even non-equilibrium curved interfaces, showing the robustness of the construction of the free energy functional.

The simulations were repeated in a quaternary system as well (see Fig. 9), where the dimensionless interfacial tensions were $\hat{\sigma}_{12}=1.0$, $\hat{\sigma}_{13}=1.1$, $\hat{\sigma}_{14}=0.75$, $\hat{\sigma}_{23}=0.9$, $\hat{\sigma}_{24}=1.25$ and $\hat{\sigma}_{34}=1.0$, while all interface thicknesses and diffusion constants were chosen to be equal, i.e., $\hat{\delta}_{ij}=\hat{D}_{ij}=1.0$. Furthermore, we chose $A_3=1.0$ to stabilize all the binary planar interfaces. The dimensionless viscosities were $x_1=x_3=1.0$, $x_2=0.5$ and $x_4=2.0$, respectively. Our experience was quite the same as in the ternary case: The system prepared in a high-energy, strongly non-equilibrium, homogeneous multi-component state undergoes phase separation, which is enhanced by the liquid flow. In the forming pattern, the bulk -- interface -- trijunction topology dominates, as expected from the free energy functional and the energy minimizing dynamics. Furthermore, the additional components vanish at evolving interfaces and trijunctions in time. The forming patterns are also quite similar in the two cases, mostly doe to the fact that we had a majority component ($c_3$ and $c_4$ in the ternary and quaternary system, respectively), in which "bubbles" of the minority phases started to form. The final (equilibrium) pattern, however, remains a question: the system has to find a configuration containing the least possible amount of interfaces and trijunctions, and representing minimum of the free energy functional. Such a configuration, nevertheless, can be a strong function of the volume fractions of the components. For example, in a binary system with a volume fraction $1/2:1/2$ 2 binary planar interfaces should form, while in a system of volume fraction $1/10:9/10$, for example, it is not energetically preferred creating such long interfaces. Instead, a bubble of the minority component forms, thus representing lower energy. In multi-component systems, the solution of the Euler-Lagrange equations can even be degenerated, i.e. it might have multiple solutions representing local minima the system can be trapped in.

Comparing Figs 8 and 9 sheds the light on an another important detail. At $t=312.5$ [panel (a) in both Figures], the ternary system is still almost homogeneous, at least compared to the ternary system, which shows a much more developed pattern. Although both systems had similar initial conditions, $A=1/2$ and $A=1$ were used in the ternary and quaternary case, respectively. This, together with Fig. 1 (c-d) give a good impression of how the triplet term works: increasing $A_3$ means increasing penalization for multi-component states (ternary \textit{and} above, as discussed in Section III. B), which forces the system to get rid of the multi-component states faster and faster. Indeed, $A_3=1$ (Fig. 9) means a stronger penalization than $A_3=1/2$ (Fig. 8), therefore, the quaternary system eliminates the high-order states.

The long time effect of $A_3$ on the evolving pattern is, however, expected to be negligible. As long as $A_3$ is roughly in the same order of magnitude as $\max[g(\mathbf{c})]$, small perturbations around binary interfaces produce small variation in the energy relative to the interfacial tension. The key is, again, that the triplet term is used solely to stabilize the binary planar interfaces, thus resulting in a strongly finite $A_3$. In contrast, in previous multi-phase/multi-component descriptions the binary interfaces are not equilibrium solutions, and the triplet term is applied to suppress the third component, which is definitely present at the binary planar interface. In these cases, the binary planar interface solution is recovered for $A_3 \to \infty$, which then may significantly affect the dynamics of the quasi-binary interfaces even if only a small amount of the third component is present. Summarizing, the purpose of applying the triplet term is essentially different in the two cases.

\section{Conclusions}

In this work we presented a generalization of the Cahn-Hilliard theory of liquid phase separation for arbitrary number of components. It has been shown that the generalization can be done on a systematic way. First a general, physically and mathematically consistent, entropy producing advection-diffusion dynamics has been set up, which then has been extended with the generalization of the Cahn-Hilliard free energy functional for many components. The extension has been done on phenomenological basis, resulting in a model, which (i) reduces/extends naturally on the level of both the free energy functional and the dynamic equations when removing/adding a component, (ii) recovers the standard  Cahn-Hilliard model for $N=2$. Furthermore, (iii) the bulk states and the two-component interfaces are \textit{stable} equilibrium solutions of the multi-component model, (iv) the free energy functional penalizes the high-order multi-component states strictly monotonously as a function of the number of components being present, and (v) the pairwise interfacial properties (interfacial tension and interface thickness) can be chosen independently.

We have shown that (i) a simple triplet energy term can be used to \textit{stabilize the binary planar interfaces}, (ii) the equilibrium contact angles are in perfect agreement with theoretical values. Furthermore, we demonstrated, that (iii) the system undergoes spinodal decomposition, when starting from a high-energy non-equilibrium state, and converges to equilibrium by developing the bulk -- interface -- trijunction topology in 2 dimensions in asymmetric ternary and quaternary systems.

Our results might significantly contribute to the continuum theory of multi-component liquids, since controlled pattern formation in these systems is  of increasing importance in several practical applications. For instance, surfactant controlled nanoshell formation opened a new chapter in targeted drug delivery \cite{nanoshell}. Another crucial field is energy: a controlled emulsion $\to$ emulsion transition in the CO$_2$/water/heavy crude oil system would result in an efficient and environmentally sound combination of CO$_2$ storage and Enhanced Oil Recovery \cite{PhysRevE.91.032404,C5CP02357B}. 

\section{Acknowledgement}

The authors thank Prof. L\'aszl\'o Gr\'an\'asy and Prof. Tam\'as Pusztai from the Wigner Research Center for Physics, Hungary, Prof. Valeriy I. Levitas from Iowa State Universiry, IA, USA, and Dr. Kumar Ankit from Karlsruhe University, Germany. This work has been supported by the VISTA basic research programme project No. 6359 "Surfactants for water/CO$_2$/hydrocarbon emulsions for combined CO$_2$ storage and utilization" of the Norwegian Academy of Science and Letters and the Statoil.

\section*{Appendix}

\subsection*{A. Energy hierarchy}

In a symmetric system, the free energy landscape reads:
\begin{equation}
\label{eq:appf}
\frac{f(\mathbf{c})}{w_0} = g(\mathbf{c}) + a\,f_3(\mathbf{c}) \enskip ,
\end{equation}
where 
\begin{equation}
g(\mathbf{c}) = \frac{1}{12} + \sum_{i=1}^N \left( \frac{c_i^4}{4}- \frac{c_i^3}{3}\right) + \frac{1}{2}\sum_{i<j} (c_i c_j)^2 \enskip , 
\end{equation}
$a=A_3/w_0 \geq 0$, and
\begin{equation}
f_3(\mathbf{c}) = \sum_{i<j<k}^{N,N,N} |c_i|\,|c_j|\,|c_k| \enskip .
\end{equation}
For $\mathbf{c}_n=\mathbb{P}[(1/n,1/n,\dots,1/n,0,0,\dots,0)]$, Eq. (\ref{eq:appf}) read as:
\begin{equation}
f(n) = \frac{1}{12}\left( 1- \frac{1}{n^2}\right) + a \left[ \frac{n(n-1)(n-2)}{6} \left( \frac{1}{n}\right)^3 \right] \enskip ,
\end{equation}
which must be monotonously increasing as a function of $n=1,2,3,\dots$. The increment for $n \to n+1$ components then reads:
\begin{equation}
f(n+1)-f(n) = \frac{1+2\,n+2\,a\,(n-1)(2+3\,n)}{12\,n^2(1+n)^2} \geq 0\enskip , 
\end{equation}
which is trivially true for $n=1$ (and $a \geq 0$), and results in
\begin{equation}
a \geq d(n)=\frac{1+2\,n}{4+2\,n+6\,n^2}
\end{equation}
for $n>1$. Since $d(n)<0$, and $\lim_{n  \to \infty}d(n) = 0$, the strictly monotonously increasing tendency of the $n$-component multiple states on the free energy landscape applies for arbitrary $A_3 \geq 0$. We note, however, that this tendency is not true for higher order triplet terms, such as $(c_i c_j c_k)^2$, for example, when $f(n)$ shows a maximum for any positive $A_3$.

\subsection*{B. Equilibrium solutions}

In the multi-component system thermodynamic equilibrium is defined by the extrema of the free energy functional. The corresponding Euler-Lagrange equations of the complete multi-component problem read:
\begin{equation}
\nabla\frac{\delta F}{\delta c_i} = \nabla \frac{\delta F}{\delta c_j} 
\end{equation}
for any $i\neq j$ pairs, $i,j=1\dots N$. The functional derivatives read:
\begin{equation}
\label{eq:funcderdef}
\frac{\delta F}{\delta c_i} = \frac{\partial f}{\partial c_i} - \nabla \frac{\partial f}{\partial \nabla c_i} \enskip ,
\end{equation}
where
\begin{equation}
f = w(\mathbf{c})\,g(\mathbf{c})+ A_3\,f_3(\mathbf{c}) + \frac{\epsilon^2(\mathbf{c})}{2}\sum_{i=1}^N (\nabla c_i)^2
\end{equation}
is the integrand of the free energy functional defined by Eq. (\ref{eq:CHfunc}). Using this in Eq. (\ref{eq:funcderdef}) yields
\begin{equation}
\begin{split}
\frac{\delta F}{\delta c_i} = &\frac{\partial w}{\partial c_i}g(\mathbf{c}) + \frac{\partial \epsilon^2}{\partial c_i} \left[ \frac{1}{2}\sum_{i=1}^N (\nabla c_i)^2 \right] + \\ & w(\mathbf{c})\frac{\partial g}{\partial c_i} + A_3 \frac{\partial f_3}{\partial c_i}- \nabla \cdot \left[ \epsilon^2(\mathbf{c})\nabla c_i\right] \enskip ,
\end{split}
\end{equation}
where
\begin{eqnarray}
\frac{\partial \epsilon^2}{\partial c_i} &=& 2\,c_i \frac{\sum_{j\neq i}[\epsilon_{ij}^2-\epsilon^2(\mathbf{c})]c_j^2}{\sum_{k<l} c_k^2 c_l^2} \\ 
\frac{\partial w}{\partial c_i} &=& 2\,c_i \frac{\sum_{j\neq i}[w_{ij}-w(\mathbf{c})]c_j^2}{\sum_{k<l} c_k^2 c_l^2} \\
\frac{\partial g}{\partial c_i} &=& c_i(\mathbf{c}^2-c_i) \\
\frac{\partial f_3}{\partial c_i} &=& \text{sgn}(c_i)\sum_{(j<k)\neq i} |c_j|\,|c_k| \enskip .
\end{eqnarray}
Since Eq. (72)-(76) vanish for $c_i(\mathbf{r})=0$, the functional derivative vanishes for a vanishing field, i.e. $(\delta F/\delta c_i)_{c_i=0}=0$. Therefore, in the binary limit $c_I(\mathbf{r})+c_J(\mathbf{r})=1$ and $c_K(\mathbf{r})=0$, the functional derivatives read:
\begin{eqnarray}
\label{eq:AEL1}\frac{\delta F}{\delta c_I} &=& w_{IJ} \frac{\partial g}{\partial c_I} - \epsilon_{IJ}^2 \nabla^2 c_I \\ 
\label{eq:AEL2}\frac{\delta F}{\delta c_J} &=& w_{IJ} \frac{\partial g}{\partial c_J} - \epsilon_{IJ}^2 \nabla^2 c_J \\ 
\frac{\delta F}{\delta c_K} &=& 0 \enskip ,
\end{eqnarray}
where $\partial g/\partial c_I = -\partial g/\partial c_J = c_I\{[c_I^2+(1-c_I)^2]-c_I\} = c_I(1-c_I)(1-2 c_I)$, i.e. $\left.\frac{\partial g}{\partial c_I}\right|_{c_I+c_J=1}=\left\{\frac{\partial}{\partial c}[c^2(1-c)^2]\right\}_{c=c_I}$. It is easy to see that the triplet term has no contribution to the free energy at all, since only 2 components are present, while $\text{sgn}(0)=0$ ensures the vanishing derivative in the equation for vanishing $c_K$. In addition, the derivatives of the Kazaryan polynomials also vanish for $c_I+c_J=1$, since in this case the sums in the nominators vanish. Substituting $c_I(x)=\{1+\tanh[x/(2\,\delta_{IJ})]\}/2$, $c_J(x)=1-c_I(x)$, and $c_K(x)=0$ into Eqns. (\ref{eq:AEL1}) and (\ref{eq:AEL2}) then yields
\begin{equation}
\delta F /\delta c_i=0
\end{equation}
for $i=1\dots N$, i.e. the binary planar interfaces are equilibrium solution of the multi-component problem.

\newpage
\bibliography{./references2}

\begin{thebibliography}{35}%
\makeatletter
\providecommand \@ifxundefined [1]{%
 \@ifx{#1\undefined}
}%
\providecommand \@ifnum [1]{%
 \ifnum #1\expandafter \@firstoftwo
 \else \expandafter \@secondoftwo
 \fi
}%
\providecommand \@ifx [1]{%
 \ifx #1\expandafter \@firstoftwo
 \else \expandafter \@secondoftwo
 \fi
}%
\providecommand \natexlab [1]{#1}%
\providecommand \enquote  [1]{``#1''}%
\providecommand \bibnamefont  [1]{#1}%
\providecommand \bibfnamefont [1]{#1}%
\providecommand \citenamefont [1]{#1}%
\providecommand \href@noop [0]{\@secondoftwo}%
\providecommand \href [0]{\begingroup \@sanitize@url \@href}%
\providecommand \@href[1]{\@@startlink{#1}\@@href}%
\providecommand \@@href[1]{\endgroup#1\@@endlink}%
\providecommand \@sanitize@url [0]{\catcode `\\12\catcode `\$12\catcode
  `\&12\catcode `\#12\catcode `\^12\catcode `\_12\catcode `\%12\relax}%
\providecommand \@@startlink[1]{}%
\providecommand \@@endlink[0]{}%
\providecommand \url  [0]{\begingroup\@sanitize@url \@url }%
\providecommand \@url [1]{\endgroup\@href {#1}{\urlprefix }}%
\providecommand \urlprefix  [0]{URL }%
\providecommand \Eprint [0]{\href }%
\providecommand \doibase [0]{http://dx.doi.org/}%
\providecommand \selectlanguage [0]{\@gobble}%
\providecommand \bibinfo  [0]{\@secondoftwo}%
\providecommand \bibfield  [0]{\@secondoftwo}%
\providecommand \translation [1]{[#1]}%
\providecommand \BibitemOpen [0]{}%
\providecommand \bibitemStop [0]{}%
\providecommand \bibitemNoStop [0]{.\EOS\space}%
\providecommand \EOS [0]{\spacefactor3000\relax}%
\providecommand \BibitemShut  [1]{\csname bibitem#1\endcsname}%
\let\auto@bib@innerbib\@empty
%</preamble>
\bibitem [{\citenamefont {Haase}\ and\ \citenamefont
  {Brujic}(2014)}]{ANGE:ANGE201406040}%
  \BibitemOpen
  \bibfield  {author} {\bibinfo {author} {\bibfnamefont {M.~F.}\ \bibnamefont
  {Haase}}\ and\ \bibinfo {author} {\bibfnamefont {J.}~\bibnamefont {Brujic}},\
  }\href {\doibase 10.1002/ange.201406040} {\bibfield  {journal} {\bibinfo
  {journal} {Angewandte Chemie}\ }\textbf {\bibinfo {volume} {126}},\ \bibinfo
  {pages} {11987} (\bibinfo {year} {2014})}\BibitemShut {NoStop}%
\bibitem [{\citenamefont {Shukutani}\ \emph {et~al.}(2014)\citenamefont
  {Shukutani}, \citenamefont {Myojo}, \citenamefont {Nakanishi}, \citenamefont
  {Norisuye},\ and\ \citenamefont {Tran-Cong-Miyata}}]{doi:10.1021/ma500302k}%
  \BibitemOpen
  \bibfield  {author} {\bibinfo {author} {\bibfnamefont {T.}~\bibnamefont
  {Shukutani}}, \bibinfo {author} {\bibfnamefont {T.}~\bibnamefont {Myojo}},
  \bibinfo {author} {\bibfnamefont {H.}~\bibnamefont {Nakanishi}}, \bibinfo
  {author} {\bibfnamefont {T.}~\bibnamefont {Norisuye}}, \ and\ \bibinfo
  {author} {\bibfnamefont {Q.}~\bibnamefont {Tran-Cong-Miyata}},\ }\href
  {\doibase 10.1021/ma500302k} {\bibfield  {journal} {\bibinfo  {journal}
  {Macromolecules}\ }\textbf {\bibinfo {volume} {47}},\ \bibinfo {pages} {4380}
  (\bibinfo {year} {2014})}\BibitemShut {NoStop}%
\bibitem [{\citenamefont {Ahearn}(1969)}]{ahearn}%
  \BibitemOpen
  \bibfield  {author} {\bibinfo {author} {\bibfnamefont {G.}~\bibnamefont
  {Ahearn}},\ }\href {\doibase 10.1007/BF02633160} {\bibfield  {journal}
  {\bibinfo  {journal} {Journal of the American Oil Chemists Society}\ }\textbf
  {\bibinfo {volume} {46}},\ \bibinfo {pages} {540A} (\bibinfo {year}
  {1969})}\BibitemShut {NoStop}%
\bibitem [{\citenamefont {Iglauer}\ \emph {et~al.}(2010)\citenamefont
  {Iglauer}, \citenamefont {Wu}, \citenamefont {Shuler}, \citenamefont {Tang},\
  and\ \citenamefont {III}}]{Iglauer201023}%
  \BibitemOpen
  \bibfield  {author} {\bibinfo {author} {\bibfnamefont {S.}~\bibnamefont
  {Iglauer}}, \bibinfo {author} {\bibfnamefont {Y.}~\bibnamefont {Wu}},
  \bibinfo {author} {\bibfnamefont {P.}~\bibnamefont {Shuler}}, \bibinfo
  {author} {\bibfnamefont {Y.}~\bibnamefont {Tang}}, \ and\ \bibinfo {author}
  {\bibfnamefont {W.~A.~G.}\ \bibnamefont {III}},\ }\href {\doibase
  http://dx.doi.org/10.1016/j.petrol.2009.12.009} {\bibfield  {journal}
  {\bibinfo  {journal} {Journal of Petroleum Science and Engineering}\ }\textbf
  {\bibinfo {volume} {71}},\ \bibinfo {pages} {23 } (\bibinfo {year}
  {2010})}\BibitemShut {NoStop}%
\bibitem [{\citenamefont {Tunio}\ \emph {et~al.}(2011)\citenamefont {Tunio},
  \citenamefont {Tunio}, \citenamefont {Ghirano},\ and\ \citenamefont
  {El~Adawy}}]{compareEOR}%
  \BibitemOpen
  \bibfield  {author} {\bibinfo {author} {\bibfnamefont {S.~Q.}\ \bibnamefont
  {Tunio}}, \bibinfo {author} {\bibfnamefont {A.~H.}\ \bibnamefont {Tunio}},
  \bibinfo {author} {\bibfnamefont {N.~A.}\ \bibnamefont {Ghirano}}, \ and\
  \bibinfo {author} {\bibfnamefont {Z.~M.}\ \bibnamefont {El~Adawy}},\
  }\href@noop {} {\bibfield  {journal} {\bibinfo  {journal} {Intl. J. Appl.
  Sci. Tech.}\ }\textbf {\bibinfo {volume} {1}},\ \bibinfo {pages} {143}
  (\bibinfo {year} {2011})}\BibitemShut {NoStop}%
\bibitem [{\citenamefont {Song}\ \emph {et~al.}(2014)\citenamefont {Song},
  \citenamefont {Li}, \citenamefont {Wei}, \citenamefont {Lai},\ and\
  \citenamefont {Bai}}]{Song201493}%
  \BibitemOpen
  \bibfield  {author} {\bibinfo {author} {\bibfnamefont {Z.}~\bibnamefont
  {Song}}, \bibinfo {author} {\bibfnamefont {Z.}~\bibnamefont {Li}}, \bibinfo
  {author} {\bibfnamefont {M.}~\bibnamefont {Wei}}, \bibinfo {author}
  {\bibfnamefont {F.}~\bibnamefont {Lai}}, \ and\ \bibinfo {author}
  {\bibfnamefont {B.}~\bibnamefont {Bai}},\ }\href {\doibase
  http://dx.doi.org/10.1016/j.compfluid.2014.03.022} {\bibfield  {journal}
  {\bibinfo  {journal} {Computers {\&} Fluids}\ }\textbf {\bibinfo {volume}
  {99}},\ \bibinfo {pages} {93 } (\bibinfo {year} {2014})}\BibitemShut
  {NoStop}%
\bibitem [{\citenamefont {Cahn}\ and\ \citenamefont
  {Hilliard}(1958)}]{:/content/aip/journal/jcp/28/2/10.1063/1.1744102}%
  \BibitemOpen
  \bibfield  {author} {\bibinfo {author} {\bibfnamefont {J.~W.}\ \bibnamefont
  {Cahn}}\ and\ \bibinfo {author} {\bibfnamefont {J.~E.}\ \bibnamefont
  {Hilliard}},\ }\href@noop {} {\bibfield  {journal} {\bibinfo  {journal} {The
  Journal of Chemical Physics}\ }\textbf {\bibinfo {volume} {28}} (\bibinfo
  {year} {1958})}\BibitemShut {NoStop}%
\bibitem [{\citenamefont {Cook}(1970)}]{Cook1970297}%
  \BibitemOpen
  \bibfield  {author} {\bibinfo {author} {\bibfnamefont {H.}~\bibnamefont
  {Cook}},\ }\href {\doibase http://dx.doi.org/10.1016/0001-6160(70)90144-6}
  {\bibfield  {journal} {\bibinfo  {journal} {Acta Metallurgica}\ }\textbf
  {\bibinfo {volume} {18}},\ \bibinfo {pages} {297 } (\bibinfo {year}
  {1970})}\BibitemShut {NoStop}%
\bibitem [{\citenamefont {Langer}(1971)}]{Langer197153}%
  \BibitemOpen
  \bibfield  {author} {\bibinfo {author} {\bibfnamefont {J.}~\bibnamefont
  {Langer}},\ }\href {\doibase http://dx.doi.org/10.1016/0003-4916(71)90162-X}
  {\bibfield  {journal} {\bibinfo  {journal} {Annals of Physics}\ }\textbf
  {\bibinfo {volume} {65}},\ \bibinfo {pages} {53 } (\bibinfo {year}
  {1971})}\BibitemShut {NoStop}%
\bibitem [{\citenamefont {Langer}(1973)}]{Langer19731649}%
  \BibitemOpen
  \bibfield  {author} {\bibinfo {author} {\bibfnamefont {J.}~\bibnamefont
  {Langer}},\ }\href {\doibase http://dx.doi.org/10.1016/0001-6160(73)90108-9}
  {\bibfield  {journal} {\bibinfo  {journal} {Acta Metallurgica}\ }\textbf
  {\bibinfo {volume} {21}},\ \bibinfo {pages} {1649 } (\bibinfo {year}
  {1973})}\BibitemShut {NoStop}%
\bibitem [{\citenamefont {Fontaine}(1972)}]{DeFontaine1972297}%
  \BibitemOpen
  \bibfield  {author} {\bibinfo {author} {\bibfnamefont {D.~D.}\ \bibnamefont
  {Fontaine}},\ }\href {\doibase
  http://dx.doi.org/10.1016/0022-3697(72)90011-X} {\bibfield  {journal}
  {\bibinfo  {journal} {Journal of Physics and Chemistry of Solids}\ }\textbf
  {\bibinfo {volume} {33}},\ \bibinfo {pages} {297 } (\bibinfo {year}
  {1972})}\BibitemShut {NoStop}%
\bibitem [{\citenamefont {Fontaine}(1973)}]{Fontaine19731285}%
  \BibitemOpen
  \bibfield  {author} {\bibinfo {author} {\bibfnamefont {D.~D.}\ \bibnamefont
  {Fontaine}},\ }\href {\doibase
  http://dx.doi.org/10.1016/S0022-3697(73)80026-5} {\bibfield  {journal}
  {\bibinfo  {journal} {Journal of Physics and Chemistry of Solids}\ }\textbf
  {\bibinfo {volume} {34}},\ \bibinfo {pages} {1285 } (\bibinfo {year}
  {1973})}\BibitemShut {NoStop}%
\bibitem [{\citenamefont {Morral}\ and\ \citenamefont
  {Cahn}(1971)}]{Morral19711037}%
  \BibitemOpen
  \bibfield  {author} {\bibinfo {author} {\bibfnamefont {J.}~\bibnamefont
  {Morral}}\ and\ \bibinfo {author} {\bibfnamefont {J.}~\bibnamefont {Cahn}},\
  }\href {\doibase http://dx.doi.org/10.1016/0001-6160(71)90036-8} {\bibfield
  {journal} {\bibinfo  {journal} {Acta Metallurgica}\ }\textbf {\bibinfo
  {volume} {19}},\ \bibinfo {pages} {1037 } (\bibinfo {year}
  {1971})}\BibitemShut {NoStop}%
\bibitem [{\citenamefont {Hoyt}(1989)}]{Hoyt19892489}%
  \BibitemOpen
  \bibfield  {author} {\bibinfo {author} {\bibfnamefont {J.}~\bibnamefont
  {Hoyt}},\ }\href {\doibase http://dx.doi.org/10.1016/0001-6160(89)90047-3}
  {\bibfield  {journal} {\bibinfo  {journal} {Acta Metallurgica}\ }\textbf
  {\bibinfo {volume} {37}},\ \bibinfo {pages} {2489 } (\bibinfo {year}
  {1989})}\BibitemShut {NoStop}%
\bibitem [{\citenamefont {Hoyt}(1990)}]{Hoyt1990227}%
  \BibitemOpen
  \bibfield  {author} {\bibinfo {author} {\bibfnamefont {J.}~\bibnamefont
  {Hoyt}},\ }\href {\doibase http://dx.doi.org/10.1016/0956-7151(90)90052-I}
  {\bibfield  {journal} {\bibinfo  {journal} {Acta Metallurgica et Materialia}\
  }\textbf {\bibinfo {volume} {38}},\ \bibinfo {pages} {227 } (\bibinfo {year}
  {1990})}\BibitemShut {NoStop}%
\bibitem [{\citenamefont {Maier-Paape}\ \emph {et~al.}(2000)\citenamefont
  {Maier-Paape}, \citenamefont {Stoth},\ and\ \citenamefont
  {Wanner}}]{stanislaus}%
  \BibitemOpen
  \bibfield  {author} {\bibinfo {author} {\bibfnamefont {S.}~\bibnamefont
  {Maier-Paape}}, \bibinfo {author} {\bibfnamefont {B.}~\bibnamefont {Stoth}},
  \ and\ \bibinfo {author} {\bibfnamefont {T.}~\bibnamefont {Wanner}},\ }\href
  {\doibase 10.1023/A:1018687811688} {\bibfield  {journal} {\bibinfo  {journal}
  {Journal of Statistical Physics}\ }\textbf {\bibinfo {volume} {98}},\
  \bibinfo {pages} {871} (\bibinfo {year} {2000})}\BibitemShut {NoStop}%
\bibitem [{\citenamefont {Korteweg}(1901)}]{Korteweg1901}%
  \BibitemOpen
  \bibfield  {author} {\bibinfo {author} {\bibfnamefont {D.~J.}\ \bibnamefont
  {Korteweg}},\ }\href@noop {} {\bibfield  {journal} {\bibinfo  {journal}
  {Arch. Neerl. Sci. Ex. Nat.}\ }\textbf {\bibinfo {volume} {6}},\ \bibinfo
  {pages} {1} (\bibinfo {year} {1901})}\BibitemShut {NoStop}%
\bibitem [{\citenamefont {Evans}(1979)}]{doi:10.1080/00018737900101365}%
  \BibitemOpen
  \bibfield  {author} {\bibinfo {author} {\bibfnamefont {R.}~\bibnamefont
  {Evans}},\ }\href {\doibase 10.1080/00018737900101365} {\bibfield  {journal}
  {\bibinfo  {journal} {Advances in Physics}\ }\textbf {\bibinfo {volume}
  {28}},\ \bibinfo {pages} {143} (\bibinfo {year} {1979})}\BibitemShut
  {NoStop}%
\bibitem [{\citenamefont
  {Salmon}(1988)}]{doi:10.1146/annurev.fl.20.010188.001301}%
  \BibitemOpen
  \bibfield  {author} {\bibinfo {author} {\bibfnamefont {R.}~\bibnamefont
  {Salmon}},\ }\href {\doibase 10.1146/annurev.fl.20.010188.001301} {\bibfield
  {journal} {\bibinfo  {journal} {Annual Review of Fluid Mechanics}\ }\textbf
  {\bibinfo {volume} {20}},\ \bibinfo {pages} {225} (\bibinfo {year}
  {1988})}\BibitemShut {NoStop}%
\bibitem [{\citenamefont {Wheeler}\ and\ \citenamefont
  {McFadden}(1997)}]{Wheeler08081997}%
  \BibitemOpen
  \bibfield  {author} {\bibinfo {author} {\bibfnamefont {A.~A.}\ \bibnamefont
  {Wheeler}}\ and\ \bibinfo {author} {\bibfnamefont {G.~B.}\ \bibnamefont
  {McFadden}},\ }\href {\doibase 10.1098/rspa.1997.0086} {\bibfield  {journal}
  {\bibinfo  {journal} {Proceedings of the Royal Society of London. Series A:
  Mathematical, Physical and Engineering Sciences}\ }\textbf {\bibinfo {volume}
  {453}},\ \bibinfo {pages} {1611} (\bibinfo {year} {1997})}\BibitemShut
  {NoStop}%
\bibitem [{\citenamefont {Anderson}\ \emph {et~al.}(2000)\citenamefont
  {Anderson}, \citenamefont {McFadden},\ and\ \citenamefont
  {Wheeler}}]{Anderson2000175}%
  \BibitemOpen
  \bibfield  {author} {\bibinfo {author} {\bibfnamefont {D.}~\bibnamefont
  {Anderson}}, \bibinfo {author} {\bibfnamefont {G.}~\bibnamefont {McFadden}},
  \ and\ \bibinfo {author} {\bibfnamefont {A.}~\bibnamefont {Wheeler}},\ }\href
  {\doibase http://dx.doi.org/10.1016/S0167-2789(99)00109-8} {\bibfield
  {journal} {\bibinfo  {journal} {Physica D: Nonlinear Phenomena}\ }\textbf
  {\bibinfo {volume} {135}},\ \bibinfo {pages} {175 } (\bibinfo {year}
  {2000})}\BibitemShut {NoStop}%
\bibitem [{\citenamefont {Anderson}\ \emph {et~al.}(1998)\citenamefont
  {Anderson}, \citenamefont {McFadden},\ and\ \citenamefont
  {Wheeler}}]{doi:10.1146/annurev.fluid.30.1.139}%
  \BibitemOpen
  \bibfield  {author} {\bibinfo {author} {\bibfnamefont {D.~M.}\ \bibnamefont
  {Anderson}}, \bibinfo {author} {\bibfnamefont {G.~B.}\ \bibnamefont
  {McFadden}}, \ and\ \bibinfo {author} {\bibfnamefont {A.~A.}\ \bibnamefont
  {Wheeler}},\ }\href {\doibase 10.1146/annurev.fluid.30.1.139} {\bibfield
  {journal} {\bibinfo  {journal} {Annual Review of Fluid Mechanics}\ }\textbf
  {\bibinfo {volume} {30}},\ \bibinfo {pages} {139} (\bibinfo {year}
  {1998})}\BibitemShut {NoStop}%
\bibitem [{\citenamefont {Tegze}\ \emph {et~al.}(2005)\citenamefont {Tegze},
  \citenamefont {Pusztai},\ and\ \citenamefont
  {Gr{\'a}n{\'a}sy}}]{Tegze2005418}%
  \BibitemOpen
  \bibfield  {author} {\bibinfo {author} {\bibfnamefont {G.}~\bibnamefont
  {Tegze}}, \bibinfo {author} {\bibfnamefont {T.}~\bibnamefont {Pusztai}}, \
  and\ \bibinfo {author} {\bibfnamefont {L.}~\bibnamefont {Gr{\'a}n{\'a}sy}},\
  }\href {\doibase http://dx.doi.org/10.1016/j.msea.2005.09.045} {\bibfield
  {journal} {\bibinfo  {journal} {Materials Science and Engineering: A}\
  }\textbf {\bibinfo {volume} {413–414}},\ \bibinfo {pages} {418 } (\bibinfo
  {year} {2005})},\ \bibinfo {note} {international Conference on Advances in
  Solidification Processes}\BibitemShut {NoStop}%
\bibitem [{\citenamefont {Kim}\ and\ \citenamefont
  {Lowengrub}(2005)}]{KimLowengrub2005}%
  \BibitemOpen
  \bibfield  {author} {\bibinfo {author} {\bibfnamefont {J.}~\bibnamefont
  {Kim}}\ and\ \bibinfo {author} {\bibfnamefont {J.}~\bibnamefont
  {Lowengrub}},\ }\href {\doibase 10.4171/IFB/132} {\bibfield  {journal}
  {\bibinfo  {journal} {Intf. Free Bound.}\ }\textbf {\bibinfo {volume} {7}},\
  \bibinfo {pages} {435 } (\bibinfo {year} {2005})}\BibitemShut {NoStop}%
\bibitem [{\citenamefont {Kim}(2012)}]{KimTernary}%
  \BibitemOpen
  \bibfield  {author} {\bibinfo {author} {\bibfnamefont {J.}~\bibnamefont
  {Kim}},\ }\href@noop {} {\bibfield  {journal} {\bibinfo  {journal} {Commun.
  Comput. Phys.}\ }\textbf {\bibinfo {volume} {12}},\ \bibinfo {pages} {613}
  (\bibinfo {year} {2012})}\BibitemShut {NoStop}%
\bibitem [{\citenamefont {T\'oth}\ \emph {et~al.}(2015)\citenamefont {T\'oth},
  \citenamefont {Pusztai},\ and\ \citenamefont
  {Gr\'an\'asy}}]{PhysRevB.92.184105}%
  \BibitemOpen
  \bibfield  {author} {\bibinfo {author} {\bibfnamefont {G.~I.}\ \bibnamefont
  {T\'oth}}, \bibinfo {author} {\bibfnamefont {T.}~\bibnamefont {Pusztai}}, \
  and\ \bibinfo {author} {\bibfnamefont {L.}~\bibnamefont {Gr\'an\'asy}},\
  }\href {\doibase 10.1103/PhysRevB.92.184105} {\bibfield  {journal} {\bibinfo
  {journal} {Phys. Rev. B}\ }\textbf {\bibinfo {volume} {92}},\ \bibinfo
  {pages} {184105} (\bibinfo {year} {2015})}\BibitemShut {NoStop}%
\bibitem [{\citenamefont {Onsager}(1945)}]{Onsager46}%
  \BibitemOpen
  \bibfield  {author} {\bibinfo {author} {\bibfnamefont {L.}~\bibnamefont
  {Onsager}},\ }\href@noop {} {\bibfield  {journal} {\bibinfo  {journal} {Ann.
  N. Y. Acad. Sci}\ }\textbf {\bibinfo {volume} {46}},\ \bibinfo {pages} {241}
  (\bibinfo {year} {1945})}\BibitemShut {NoStop}%
\bibitem [{\citenamefont {Bollada}\ \emph {et~al.}(2012)\citenamefont
  {Bollada}, \citenamefont {Jimack},\ and\ \citenamefont
  {Mullis}}]{Bollada2012816}%
  \BibitemOpen
  \bibfield  {author} {\bibinfo {author} {\bibfnamefont {P.}~\bibnamefont
  {Bollada}}, \bibinfo {author} {\bibfnamefont {P.}~\bibnamefont {Jimack}}, \
  and\ \bibinfo {author} {\bibfnamefont {A.}~\bibnamefont {Mullis}},\ }\href
  {\doibase http://dx.doi.org/10.1016/j.physd.2012.01.006} {\bibfield
  {journal} {\bibinfo  {journal} {Physica D: Nonlinear Phenomena}\ }\textbf
  {\bibinfo {volume} {241}},\ \bibinfo {pages} {816 } (\bibinfo {year}
  {2012})}\BibitemShut {NoStop}%
\bibitem [{\citenamefont {Kazaryan}\ \emph {et~al.}(2000)\citenamefont
  {Kazaryan}, \citenamefont {Wang}, \citenamefont {Dregia},\ and\ \citenamefont
  {Patton}}]{PhysRevB.61.14275}%
  \BibitemOpen
  \bibfield  {author} {\bibinfo {author} {\bibfnamefont {A.}~\bibnamefont
  {Kazaryan}}, \bibinfo {author} {\bibfnamefont {Y.}~\bibnamefont {Wang}},
  \bibinfo {author} {\bibfnamefont {S.~A.}\ \bibnamefont {Dregia}}, \ and\
  \bibinfo {author} {\bibfnamefont {B.~R.}\ \bibnamefont {Patton}},\ }\href
  {\doibase 10.1103/PhysRevB.61.14275} {\bibfield  {journal} {\bibinfo
  {journal} {Phys. Rev. B}\ }\textbf {\bibinfo {volume} {61}},\ \bibinfo
  {pages} {14275} (\bibinfo {year} {2000})}\BibitemShut {NoStop}%
\bibitem [{\citenamefont {Ankit}\ \emph {et~al.}(2013)\citenamefont {Ankit},
  \citenamefont {Nestler}, \citenamefont {Selzer},\ and\ \citenamefont
  {Reichardt}}]{kumarbritta}%
  \BibitemOpen
  \bibfield  {author} {\bibinfo {author} {\bibfnamefont {K.}~\bibnamefont
  {Ankit}}, \bibinfo {author} {\bibfnamefont {B.}~\bibnamefont {Nestler}},
  \bibinfo {author} {\bibfnamefont {M.}~\bibnamefont {Selzer}}, \ and\ \bibinfo
  {author} {\bibfnamefont {M.}~\bibnamefont {Reichardt}},\ }\href {\doibase
  10.1007/s00410-013-0950-x} {\bibfield  {journal} {\bibinfo  {journal}
  {Contributions to Mineralogy and Petrology}\ }\textbf {\bibinfo {volume}
  {166}},\ \bibinfo {pages} {1709} (\bibinfo {year} {2013})}\BibitemShut
  {NoStop}%
\bibitem [{\citenamefont {Levitas}\ and\ \citenamefont
  {Roy}(2015)}]{PhysRevB.91.174109}%
  \BibitemOpen
  \bibfield  {author} {\bibinfo {author} {\bibfnamefont {V.~I.}\ \bibnamefont
  {Levitas}}\ and\ \bibinfo {author} {\bibfnamefont {A.~M.}\ \bibnamefont
  {Roy}},\ }\href {\doibase 10.1103/PhysRevB.91.174109} {\bibfield  {journal}
  {\bibinfo  {journal} {Phys. Rev. B}\ }\textbf {\bibinfo {volume} {91}},\
  \bibinfo {pages} {174109} (\bibinfo {year} {2015})}\BibitemShut {NoStop}%
\bibitem [{\citenamefont {Tegze}\ \emph {et~al.}(2009)\citenamefont {Tegze},
  \citenamefont {Bansel}, \citenamefont {T\'oth}, \citenamefont {Pusztai},
  \citenamefont {Fan},\ and\ \citenamefont {Gr\'an\'asy}}]{Tegze20091612}%
  \BibitemOpen
  \bibfield  {author} {\bibinfo {author} {\bibfnamefont {G.}~\bibnamefont
  {Tegze}}, \bibinfo {author} {\bibfnamefont {G.}~\bibnamefont {Bansel}},
  \bibinfo {author} {\bibfnamefont {G.~I.}\ \bibnamefont {T\'oth}}, \bibinfo
  {author} {\bibfnamefont {T.}~\bibnamefont {Pusztai}}, \bibinfo {author}
  {\bibfnamefont {Z.}~\bibnamefont {Fan}}, \ and\ \bibinfo {author}
  {\bibfnamefont {L.}~\bibnamefont {Gr\'an\'asy}},\ }\href {\doibase
  http://dx.doi.org/10.1016/j.jcp.2008.11.011} {\bibfield  {journal} {\bibinfo
  {journal} {Journal of Computational Physics}\ }\textbf {\bibinfo {volume}
  {228}},\ \bibinfo {pages} {1612 } (\bibinfo {year} {2009})}\BibitemShut
  {NoStop}%
\bibitem [{\citenamefont {De~Villiers}\ and\ \citenamefont
  {Lvov}(2007)}]{nanoshell}%
  \BibitemOpen
  \bibfield  {author} {\bibinfo {author} {\bibfnamefont {M.~M.}\ \bibnamefont
  {De~Villiers}}\ and\ \bibinfo {author} {\bibfnamefont {Y.~M.}\ \bibnamefont
  {Lvov}},\ }\enquote {\bibinfo {title} {Nanoshells for drug delivery},}\ in\
  \href {\doibase 10.1002/9783527610419.ntls0118} {\emph {\bibinfo {booktitle}
  {Nanotechnologies for the Life Sciences}}}\ (\bibinfo  {publisher} {Wiley-VCH
  Verlag GmbH {\&} Co. KGaA},\ \bibinfo {year} {2007})\BibitemShut {NoStop}%
\bibitem [{\citenamefont {T\'oth}\ and\ \citenamefont
  {Kvamme}(2015)}]{PhysRevE.91.032404}%
  \BibitemOpen
  \bibfield  {author} {\bibinfo {author} {\bibfnamefont {G.~I.}\ \bibnamefont
  {T\'oth}}\ and\ \bibinfo {author} {\bibfnamefont {B.}~\bibnamefont
  {Kvamme}},\ }\href {\doibase 10.1103/PhysRevE.91.032404} {\bibfield
  {journal} {\bibinfo  {journal} {Phys. Rev. E}\ }\textbf {\bibinfo {volume}
  {91}},\ \bibinfo {pages} {032404} (\bibinfo {year} {2015})}\BibitemShut
  {NoStop}%
\bibitem [{\citenamefont {T{\'o}th}\ and\ \citenamefont
  {Kvamme}(2015)}]{C5CP02357B}%
  \BibitemOpen
  \bibfield  {author} {\bibinfo {author} {\bibfnamefont {G.~I.}\ \bibnamefont
  {T{\'o}th}}\ and\ \bibinfo {author} {\bibfnamefont {B.}~\bibnamefont
  {Kvamme}},\ }\href {\doibase 10.1039/C5CP02357B} {\bibfield  {journal}
  {\bibinfo  {journal} {Phys. Chem. Chem. Phys.}\ }\textbf {\bibinfo {volume}
  {17}},\ \bibinfo {pages} {20259} (\bibinfo {year} {2015})}\BibitemShut
  {NoStop}%
\end{thebibliography}%

\end{document}